\documentclass[aps,prl,superscriptaddress,reprint,floats,citeautoscript,floatfix,nobibnotes,nofootinbib]{revtex4-1}

\usepackage[intlimits,tbtags]{amsmath}
\usepackage{times}
\usepackage{color}
\usepackage{amssymb}
\usepackage{graphicx}
\usepackage{bm}
\usepackage{array}
\usepackage{dcolumn}
\usepackage{bm}
\usepackage{setspace}

\newcolumntype{d}[1]{D{.}{.}{#1} }
\usepackage{xcolor}

\newcommand{\Tc}{$T_\text{c}$}

\newcommand{\jsnu}{Laboratory of Quantum Functional Materials Design and Application, School of Physics and Electronic Engineering, Jiangsu Normal University, Xuzhou 221116, China}

\newcommand{\poland}{Institute of Physics, Cz\c{e}stochowa University of Technology, Ave. Armii Krajowej 19, 42-200 Cz\c{e}stochowa, Poland}
\newcommand{\jilin}{Key Laboratory of Material Simulation Methods and Software of Ministry of Education
and State Key Laboratory of Superhard Materials, College of Physics, Jilin University, Changchun 130012, China}

\begin{document}

\title { Prediction of high-\Tc\ superconductivity under submegabar pressure in ternary actinium borohydrides }
\author{Tingting Gu}
\affiliation{\jsnu}
\author{Wenwen Cui}\email{wenwencui@jsnu.edu.cn}
\author{Jian Hao}
\author{Jingming Shi}
\affiliation{\jsnu}
\author{Artur P. Durajski}
\affiliation{\poland}
\author{Hanyu Liu}\email{hanyuliu@jlu.edu.cn}
\affiliation{\jilin}
\author{Yinwei Li}\email{yinwei\_li@jsnu.edu.cn}
\affiliation{\jsnu}


\begin{abstract}
  Ternary hydrides are considered as the idea candidates with high critical temperature (\Tc) stabilized at submegabar pressure, evidenced by the  recent discoveries in LaBeH$_8$ (110 K at 80 GPa ) and LaB$_2$H$_8$ (106 K at 90 GPa). Here, we investigate the crystal structures and superconductivity of Ac-B-H system under pressures of 100 and 200 GPa by using an advanced structure method combined with first-principles calculations. As a result, nine stable compounds were identified, where B atoms are bonded with  H atoms in formation with diverse BH$_x$ motifs, e.g., methanelike (BH$_4$), polythenelike, (BH$_2$)$_n$, and BH$_6$ octahedron. Among them, seven Ac-B-H compounds were found to become superconductive. In particular, AcBH$_7$ was estimated to have a \Tc\ of 122 K at 70 GPa. Our in-depth analysis reveals that the B-H interactions in the BH$_6$ units play a key role in its high superconductivity and stability at submegar pressure. Our current results provide a guidance for future experiments to synthesize ternary hydride superconductors with high-\Tc\ at moderate pressure.

\end{abstract}
\maketitle

\section{INTRODUCTION}
 Hydrogen is considered as the promising candidate for high-temperature superconductor, due to small atomic mass and strong phonon vibration frequency~\cite{wigner1935possibility,ashcroft1968metallic}. However, the metallization pressure required to compress hydrogen exceeds 450 GPa, which is difficult to achieve via current hydrostatic pressure experiments~\cite{azadi2014dissociation,mcminis2015molecular}. Ashcroft pioneered  the possibility of finding high-temperature superconductors in hydrides because of the 'chemical pre-compression' effect of non-hydrogen elements, which allows for the metallization of hydrides at lower pressures~\cite{ashcroft2004hydrogen} than that for pure hydrogen. However, to search for a suitable hydride with high superconductivity remained a great challenge at that time. Until recent decades, thanks to the rapid development of computational structure prediction algorithms, many hydrides were predicted to become stable at megabar pressures with high superconducting critical temperature above 200 K. These theoretical studies have led to a surge in the number of high-temperature superconductors synthesized in binary hydrides, e.g., H$_3$S~\cite{li2014metallization,duan2015pressure,duan2014pressure,drozdov2015conventional}, LaH$_{10}$~\cite{liu2017potential,drozdov2019superconductivity,peng2017hydrogen,somayazulu2019evidence}, YH$_6$~\cite{li2015pressure,kong2021superconductivity,troyan2021anomalous}, YH$_9$~\cite{kong2021superconductivity,snider2021synthesis} and CaH$_6$~\cite{wang2012superconductive,ma2022high,li2022superconductivity1}.

 Almost all possibilities have been theoretically explored for binary hydrides~\cite{zurek2019high,flores2020perspective}, thus it is highly desirable  to explore ternary hydrides considering the addition of new elements and new configurations, which expands the material field to find structures in the ternary hydrides  with higher \Tc\ at relatively lower pressures. To date, increasing amounts of ternary hydrides have been synthesized experimentally with the high-temperature superconductivity: (La,Y)H$_{10}$ (253 K at 183 GPa)~\cite{semenok2021superconductivity}, (La,Ce)H$_9$ (178 K at 110 GPa or 176 K at 100 GPa)~\cite{chen2022enhancement,bi2022efficient}, (La,Nd)H$_{10}$ (148 K at 170 GPa)~\cite{semenok2022effect} and (La,Al)H$_{10}$ (223 K at 164 GPa)~\cite{chen2024high}. However, the pressure still exceeds 100 GPa beyond further application. Recently, researches proposed an effective strategy  to reduce the external pressure by adding a light element that binds with H atoms to form small molecules, acting as pre-compression factor, such as CH$_4$ in CSH$_7 $~\cite{cui2020route} and MC$_2$H$_8$~\cite{jiang2022high}, B/BeH$_8$ in X(B/Be)H$_8$ ~\cite{di2021bh,liang2021prediction,belli2022impact,song2023stoichiometric,gao2024prediction}, and BH$_4$ in XB$_2$H$_8$~\cite{gao2021phonon,li2022superconductivity2}. In this way, the newly constructed structures could not only become stable at lower pressure but also remain good superconductivity. Recently, structures with Mg$_2$XH$_6$-type are proposed to be an ideal superconductor at ambient pressure, e.g., Mg$_2$IrH$_6$ (\Tc\ = 160 K) ~\cite{sanna2024prediction,dolui2024feasible}. Particularly, LaBeH$_8$ and LaB$_2$H$_8$ have been synthesized successfully below megabar pressure with \Tc\ of 110 K at 80 GPa~\cite{song2023stoichiometric} and 106 K at 90 GPa~\cite{song2024superconductivity}, respectively, which greatly motivated us to hunt for high-\Tc\ superconductors at moderate pressures in ternary hydrides.

 The $p$$^{0}$ and $d$$^{1}$ metals with low-lying empty orbitals tend to form phonon-mediated high-\Tc\ superconducting (HTSC) metal polyhydrides. Among them, actinium, which is similar to lanthanum in chemical properties, atomic size, electronegativity, and electronic configuration, forms many actinium hydrides with good superconducting properties: $R\bar3m$ AcH$_{10}$ (251 K at 200 GPa), $I$4/$mmm$ AcH$_{12}$ (173 K at 150 GPa) and $P\bar6m2$ AcH$_{16}$ (241 K at 150 GPa)~\cite{semenok2018actinium}.
 In addition, a large number of ternary borohydrides  have been explored, e.g., Li-B-H~\cite{kokail2017prediction}, S-B-H~\cite{du2019phase}, Ca-B-H~\cite{di2020phase}, Na-B-H~\cite{li2022structural}, Mg-B-H~\cite{sukmas2023first} and  La-B-H in particular, with \Tc\ of 126 K at 50 GPa~\cite{di2021bh,liang2021prediction,belli2022impact}. On the basis, it is highly required to explore the crystal structure and superconductivity of the Ac-B-H system  in anticipation of finding high-temperature superconductors stabilized at low pressures. In this work, the crystal structure, stability and superconducting properties of Ac$_x$B$_y$H$_z$ ($x$ = 1-2, $y$ = 1-2, $z$ = 1-16) at 100 GPa and 200 GPa are theoretically investigated. Nine thermodynamically stable compounds are uncovered, namely, $P$6/$mmm$ AcBH, $P$2$_1$/$m$ AcBH$_4$, $P$2$_1$/$m$ AcB$_2$H$_7$, $Pbnm$ AcBH$_6$, $Pnm$2$_1$ AcB$_2$H$_{14}$, $Pm$2$_1$$b$ AcB$_2$H$_{13}$, $Pm$2$_1$$b$ AcBH$_{16}$, $P\bar3m1$ AcBH$_7$ and $P$2$_1$/$m$ AcBH$_8$. In particularly, $P\bar3m1$ AcBH$_7$ remain dynamically stable at 70 GPa with \Tc\ of 122 K, comparable to that of LaBeH$_8$ (110 K at 80 GPa)~\cite{song2023stoichiometric} and LaB$_2$H$_8$ (106 K at 90 GPa)~\cite{song2024superconductivity}, which have already been synthesized experimentally.

\raggedbottom
\begin{figure*}[htp]
\centering
  \includegraphics[width=0.8\linewidth,angle=0]{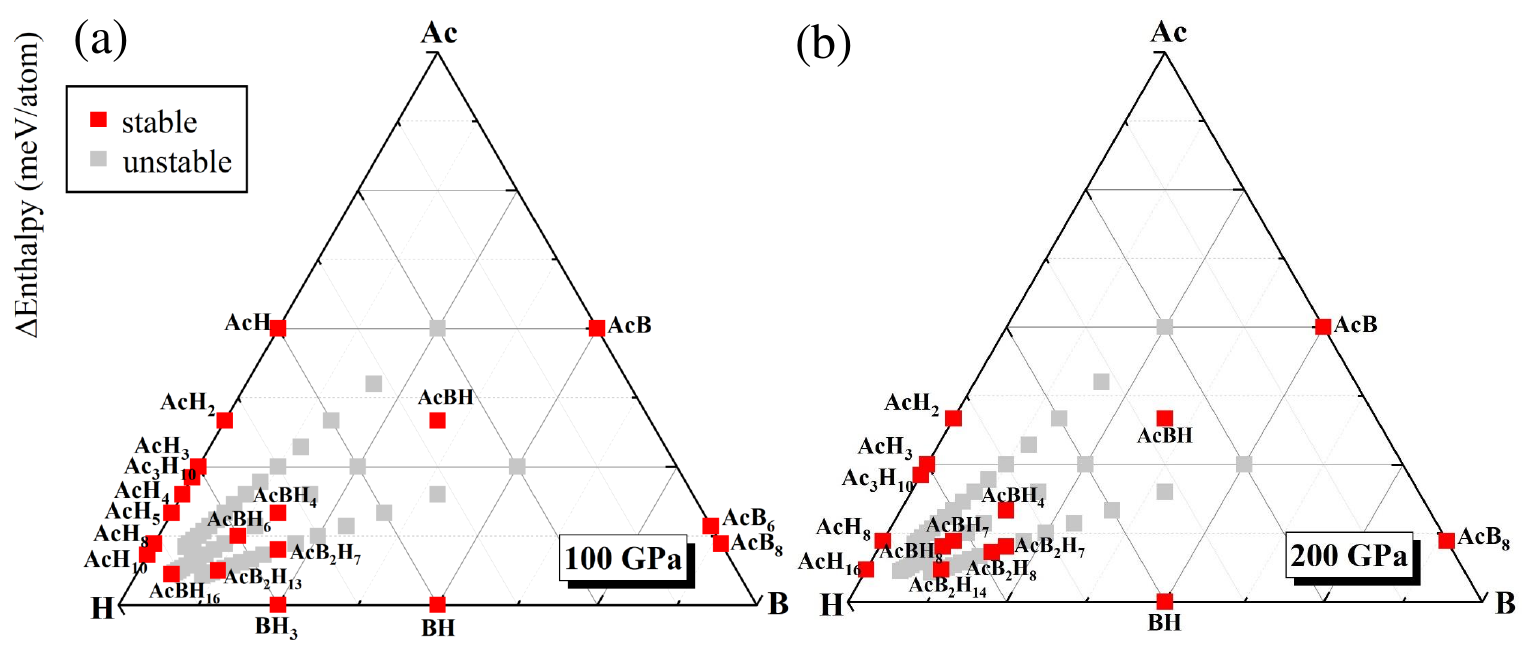}
  \caption{\label{fig:1} Calculated stabilities of Ac$_x$B$_y$H$_z$ relative to Ac, B, H, and binary compounds at (a) 100~GPa and (b) 200~GPa. Red and grey squares denote the thermodynamically stable and ternary stoichiometries, respectively.}
\end{figure*}

\section{Computational Details}

 The candidate structures of Ac-B and Ac-B-H systems are predicted using the {\sc calypso} code~\cite{wang2010crystal,wang2012calypso,gao2019interface,shao2022symmetry} based on a particle swarm optimisation algorithm. We perform composition crystal structure searches for Ac$_x$B$_y$ ($x$ = 1-2, $y$ = 1-10) and Ac$_x$B$_y$H$_z$ ($x$ = 1-2, $y$ = 1-2, $z$ = 1-16) ranging from 1 to 4 f.u./cell at 100 and 200 GPa. More than 2000 structures for each stoichiometry during the prediction search and continues to generate 1000 structures after the lowest energy structure is determined in order to ensure the convergence. The phonon calculations were performed for all structures by using a supercell method as implemented in {\sc phonopy} code~\cite{parlinski1997first,togo2008first}. The total energy and electronic properties are calculated by using the Vienna $Ab$ $initio$ Simulation Package ({\sc vasp})~\cite{kresse1996efficient} with the Perdew-Burke-Ernzerhof~\cite{perdew1996generalized,perdew1992atoms} exchange-correlation functional. Projector augmented wave (PAW) potentials with the valence electrons 6$s^2$6$p^6$6$d^1$7$s^2$ for Ac, 5$s^2$6$s^2$5$p^6$5$d^1$ for La, 5$s^2$6$s^2$5$p^6$5$d^1$4$f^1$ for Ce, 5$s^2$6$s^2$5$p^6$5$d^1$4$f^1$ for Th, 2$s^2$2$p^1$ for B, and 1$s^1$ for H are adopted~\cite{kresse1999ultrasoft}. To ensure the convergence of force and energy, we set the corresponding plane wave cutoff energy to 600~eV and Brillouin zone samplings to 0.20~\AA$^{-1}$. To consider the chemical bonding state of the structure, we have used Electron Localization Function (ELF) to analyze the covalent bonds and Bader to analyse the charge transfer~\cite{becke1990simple,bader1985atoms}. The electron-phonon coupling (EPC) constant was calculated within the framework of the linear-response theory as carried out in the {\sc quantum espresso} package~\cite{giannozzi2009quantum}. Ultra-soft pseudopotentials were used with a kinetic energy cutoff of 80 Ry. The detailed $k$ meshes, and $q$ points for superconducting Ac-B-H compounds can be found in the Supplementary Materials, SM~\cite{suppe}. 
 Both EPC parameters and \Tc\ calculated in the Ac-B-H system are based on the broadening parameter of 0.04 Rydberg. The Allen-Dynes modified McMillan (ADM) equation~\cite{allen1975transition}, which better describes superconductors with $\lambda$ $<$ 1.5, is applied to calculating it.
\begin{equation}
  T_\text{c} = \frac{\omega_{\text{log}}}{1.2} \text{exp}\left[-\frac{1.04(1+\lambda)}{\lambda-\mu^*(1+0.62\lambda)}\right],
  \end{equation}
where $\omega_{\text{log}}$ is the logarithmic average frequency, $\lambda$ is the electron phonon coupling parameter and $\mu^*$ is the Coulomb pseudopotential, often assumed to be between 0.1--0.13.

\raggedbottom
\begin{figure*}[htp]
\centering
  \includegraphics[width=0.618\linewidth,angle=0]{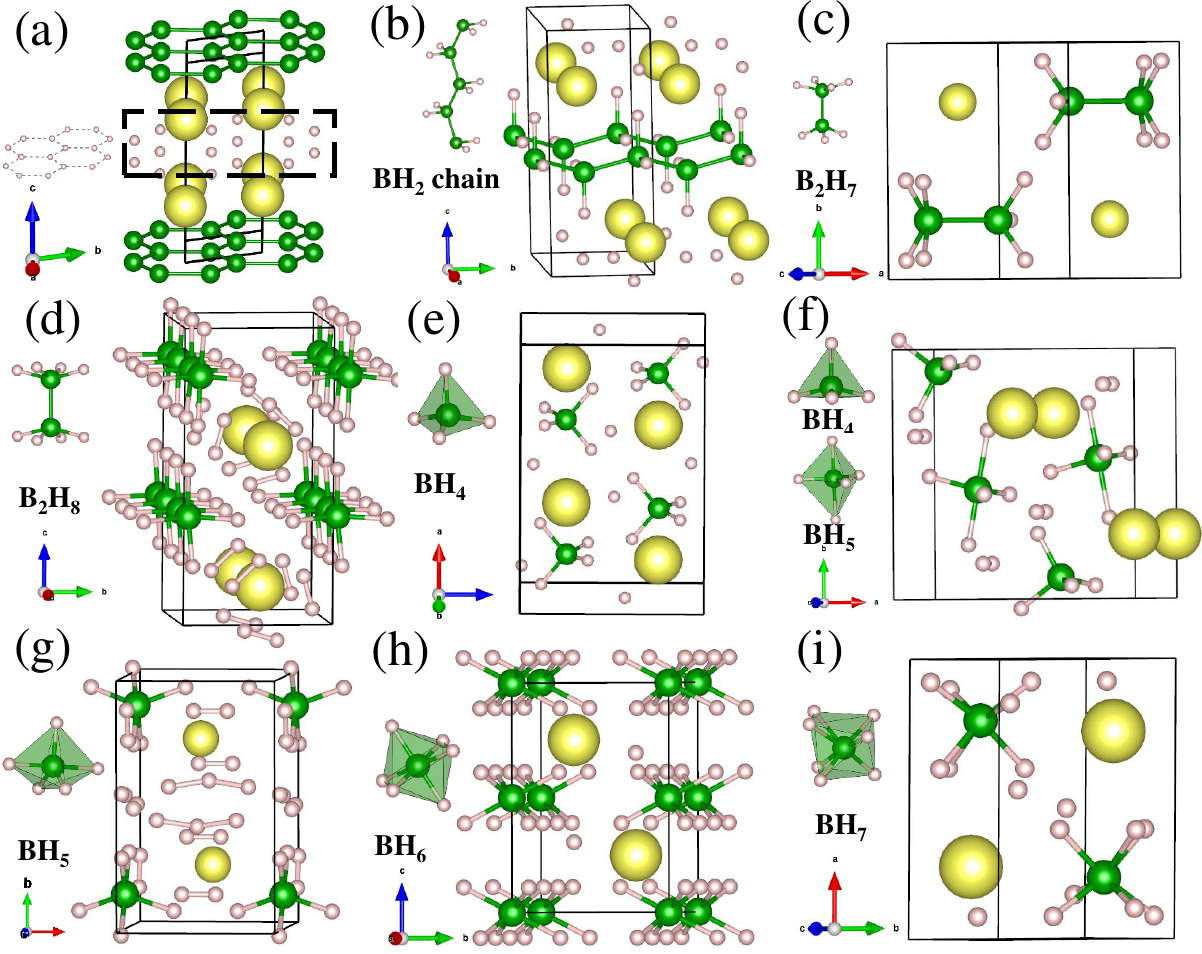}
  \caption{\label{fig:2} The stable structures of Ac-B-H compounds: (a) $P$6/$mmm$ AcBH, (b) $P$2$_1$/$m$ AcBH$_4$, (c) $P$2$_1$/$m$ AcB$_2$H$_7$, (d) $Pnm$2$_1$ AcB$_2$H$_{14}$, (e) $Pbnm$ AcBH$_6$, (f) $Pm$2$_1$$b$ AcB$_2$H$_{13}$, (g) $Pm$2$_1$$b$ AcBH$_{16}$, (h) $P\bar3m1$ AcBH$_7$ and (i) $P$2$_1$/$m$ AcBH$_8$. Yellow, Green and pink spheres represent Ac, B, and H atoms, respectively.}
\end{figure*}

\raggedbottom
\begin{figure*}[htp]
\centering
  \includegraphics[width=0.85\linewidth,angle=0]{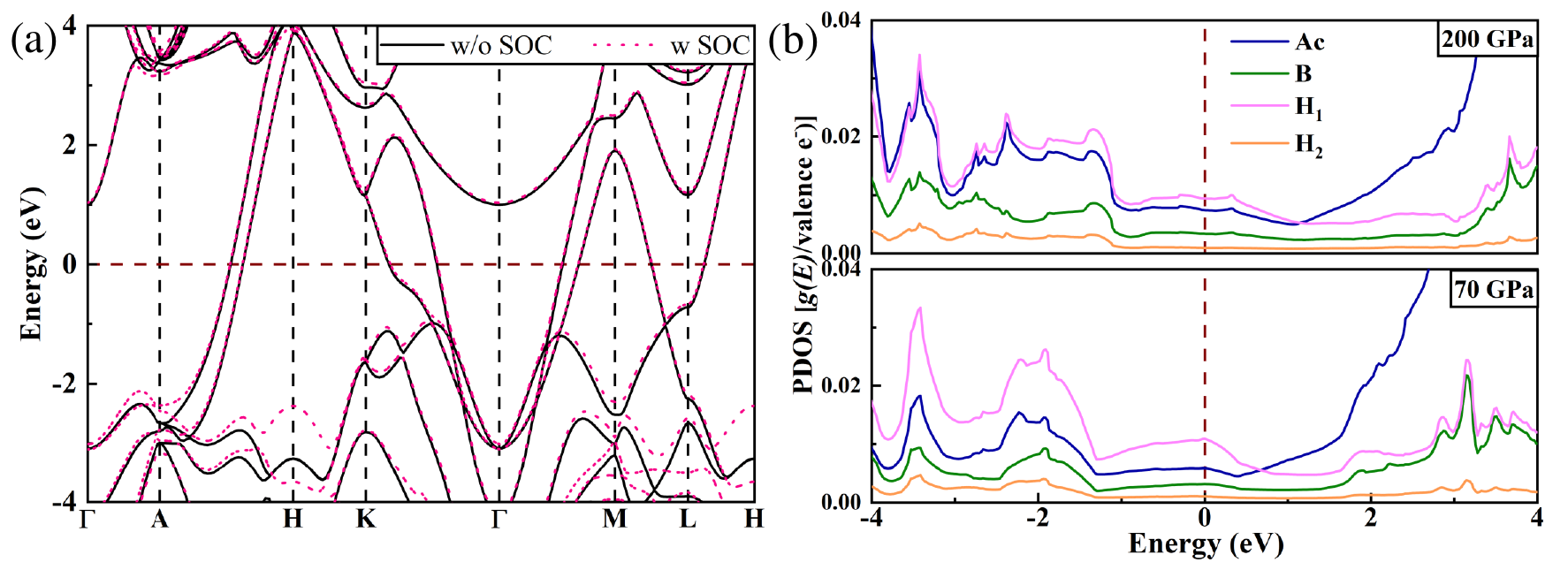}
  \caption{\label{fig:3} (a) Electronic band structures of $P\bar3m1$ AcBH$_7$ at 200~GPa without (w{/}o SOC)~
  and with the inclusion of spin-orbit coupling (w SOC). The Fermi energy is set to zero. (b) Atom-projected density of states near the Fermi level for $P\bar3m1$ AcBH$_7$ at 200 GPa and 70 GPa. H$_1$ denotes the H atoms in the BH$_6$ units, and H$_2$ denotes the H atom that is not bonded to the B atom.}
\end{figure*}

\raggedbottom
\begin{figure*}[htp]
\centering
  \includegraphics[width=0.85\linewidth,angle=0]{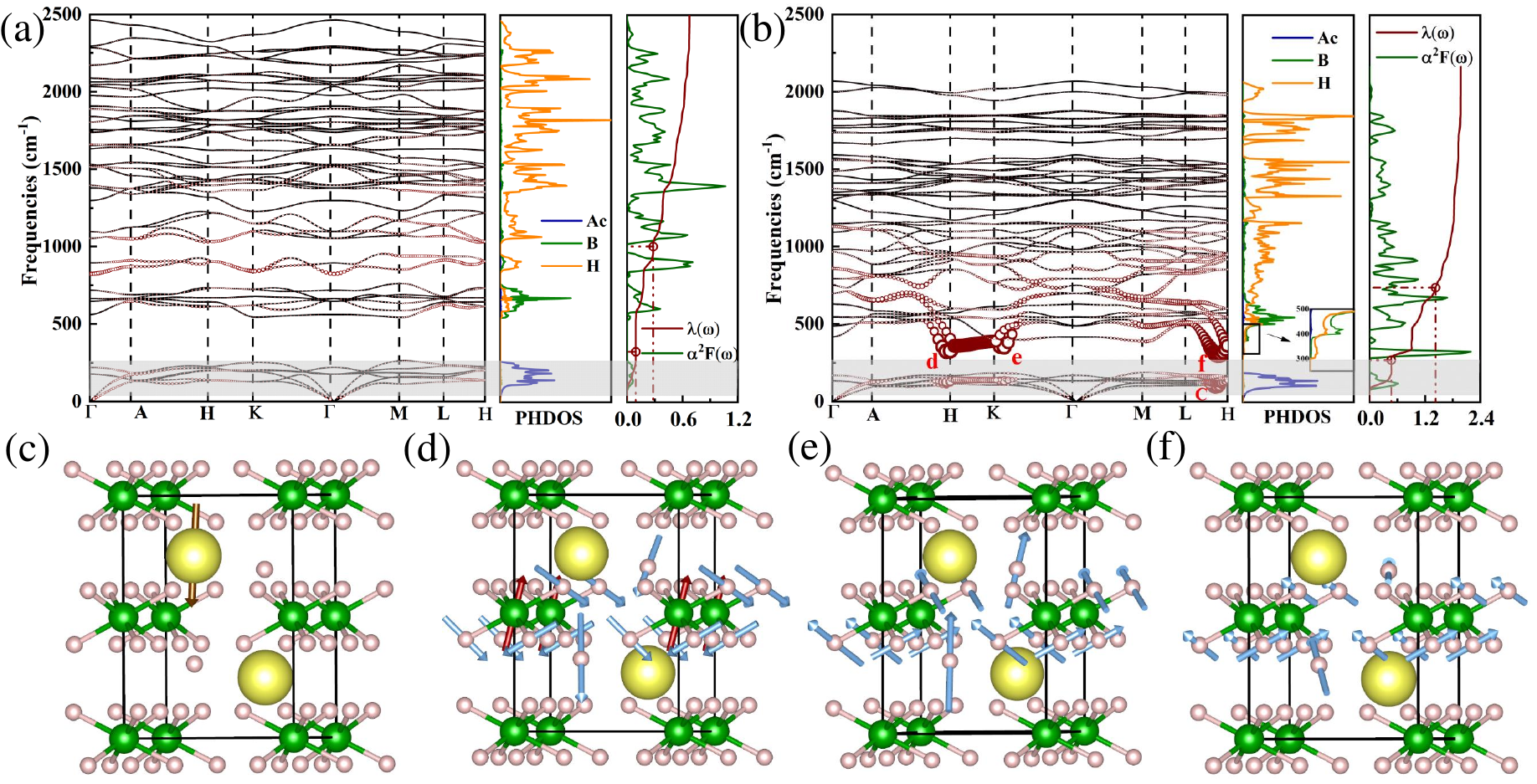}
  \caption{\label{fig:4} Phonon-dispersion curves, PHDOS, projected on the Ac, B, and H atoms, Eliashberg spectral function,$\alpha^2F(\omega)$ and $\lambda(\omega)$ for $P\bar3m1$ AcBH$_7$ at (a) 200~GPa and (b) 70~GPa. (c-f) The vibrational modes at `c', `d', `e' and `f' of $P\bar3m1$ AcBH$_7$ at 70 GPa. The direction of vibration of Ac, B and H are marked as golden, red, and blue arrows, respectively.}
\end{figure*}

\begin{table*}[t]
\centering
\renewcommand{\arraystretch}{1.2}
\setlength{\tabcolsep}{6mm}{
\caption{The calculated electron-phonon coupling parameter ($\lambda$), logarithmic average phonon frequency ($\omega_\text{log}$ (K)), and the estimated \Tc\ for $Pm$2$_1$$b$ AcBH$_{16}$, $P\bar3m1$ AcBH$_7$, $P$2$_1$/$m$ AcBH$_8$, $Pm$2$_1$$b$ AcB$_2$H$_{13}$, $Pnm$2$_1$ AcB$_2$H$_{14}$ at high pressures.}
\begin{tabular}{@{}cccccc@{}}
\hline
\hline
\multicolumn{4}{c}{}                                                   & \multicolumn{2}{c}{\Tc\ (K)} \\ \cline{5-6}
Phases                         & Pressure (GPa)    & $\lambda$ & $\omega_\text{log}$ (K)  & ADM       & Eliashberg     \\ \hline
$Pm$2$_1$$b$ AcBH$_{16}$       & 100               & 0.71        & 950.77                 & 35        & 39             \\
$P\bar3m1$ AcBH$_7$            & 200               & 0.67        & 1309.04                & 41        & 43             \\
                               & 70                & 1.98        & 600.86                 & 86        & 122            \\
$P$2$_1$/$m$ AcBH$_8$          & 200               & 0.64        & 1368.91                & 37        & 43             \\
                               & 100               & 0.71        & 1118.08                & 40        & 46             \\
$Pnm$2$_1$ AcB$_2$H$_{14}$     & 200               & 0.72        & 1108.48                & 41        & 42             \\
\hline
\hline
\end{tabular}}
\end{table*}

\section{RESULTS AND DISCUSSION}

 In order to acquire the phase diagram of ternary Ac-B-H system, we need to first determine the structures of associated binary system. The B-H system~\cite{yao2011bh3,abe2011crystalline,hu2013pressure,torabi2013pressure,murcia2018effects,yang2019novel} and the Ac-H system~\cite{semenok2018actinium} have been thoroughly investigated, while the information of Ac-B system is lacking. Thus we make a structural prediction for Ac-B at pressures of 100 and 200 GPa and identified three thermodynamically stable compounds, namely, $R\bar3m$ AcB, $Pm\bar3m$ AcB$_6$ and $R\bar3m$ AcB$_8$ (see Figs. S1-S5 in SM~\cite{suppe} for more details). The calculated phase diagram of Ac$_x$B$_y$H$_z$ at pressures of 100 and 200 GPa as shown in Fig.~\ref{fig:1}. Six compounds are identified to be thermodynamically stable at 100 GPa (e.g., $P$6/$mmm$ AcBH, $P$2$_1$/$m$ AcBH$_4$, $Pbnm$ AcBH$_6$, $Pm$2$_1$$b$ AcBH$_{16}$, $P$2$_1$/$m$ AcB$_2$H$_7$, $Pm$2$_1$$b$ AcB$_2$H$_{13}$). Three compounds AcBH, AcBH$_4$ and AcB$_2$H$_7$ remain thermodynamically stable without phase transitions as pressure increases to 200 GPa. Additionally, another four stoichiometries AcBH$_7$, AcBH$_8$, AcB$_2$H$_8$ and AcB$_2$H$_{14}$ start to locate at the convex hull at 200 GPa. Note that AcBH$_4$, AcBH$_6$, AcBH$_8$ and AcB$_2$H$_8$ have been reported in previous work~\cite{li2023pressure}, however, our proposed structures are more energetically stable with exception of AcB$_2$H$_8$ which adopts the same symmetry $I$4/$mmm$ (see Fig. S6).

\raggedbottom
\begin{figure*}[htp]
\centering
  \includegraphics[width=0.8\linewidth,angle=0]{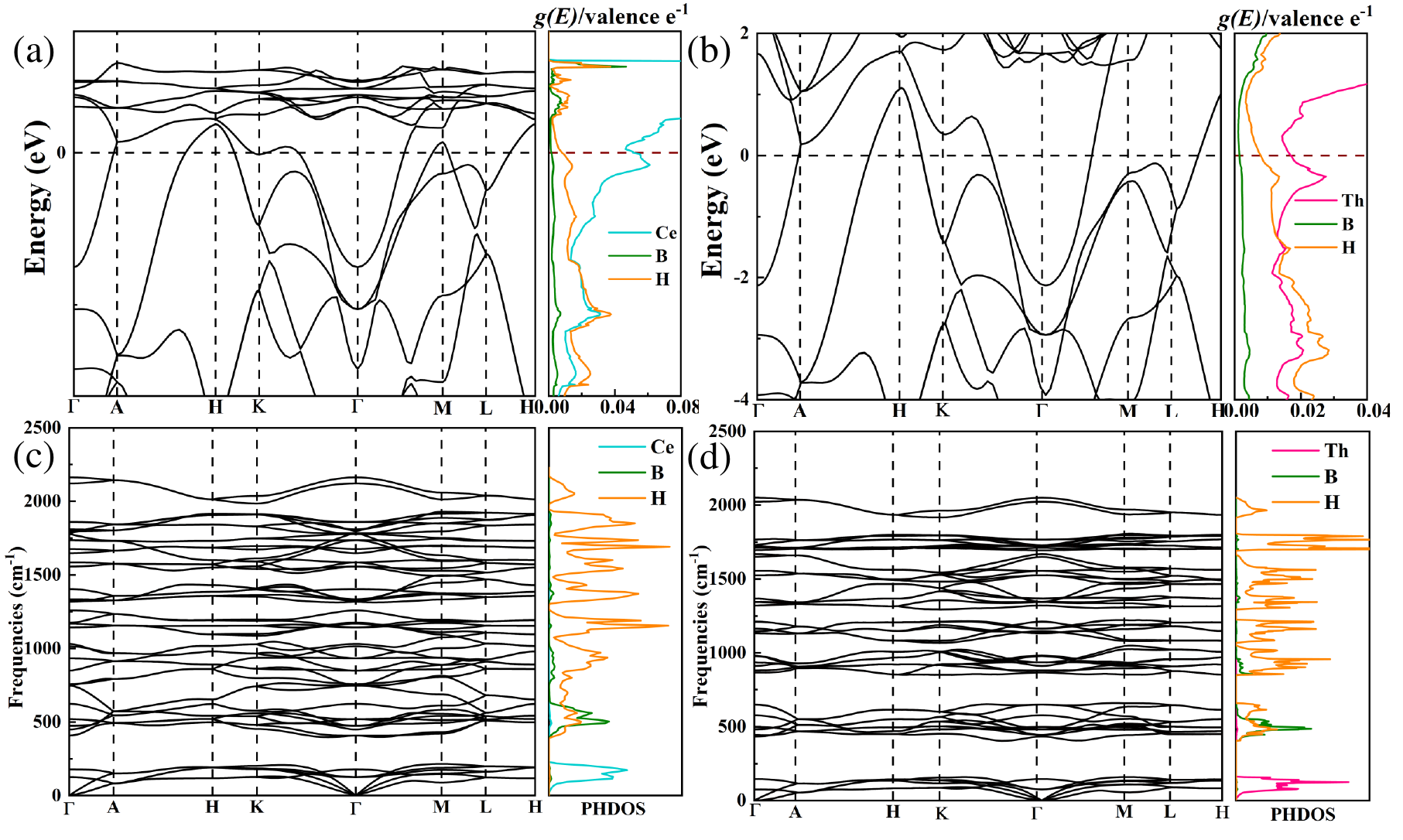}
  \caption{\label{fig:5} The projected electronic band structure of each atoms in (a) CeBH$_7$ at 75 GPa and (b) ThBH$_7$ at 50 GPa. Phonon-dispersion curves, phonon density of states (PHDOS) for (c) CeBH$_7$ at 75 GPa and (d) ThBH$_7$ at 50 GPa.}
\end{figure*}

 Figure.~\ref{fig:2} shows the structures of nine  Ac-B-H compounds with corresponding structural parameters as indicated in SM~\cite{suppe}. AcBH adopts hexagonal $P$6/$mmm$ symmetry, comprising alternating Ac, B and H layers, where B and H atoms form the honeycomb sublattice [Fig.~\ref{fig:2}(a)]. In AcBH$_4$ [Fig.~\ref{fig:2}(b)], the zigzag B covalently bonded with two hydrogen atoms to form BH$_2$ units, which are polymerized along $y$ axis and behave like polythene. For H-richer AcB$_2$H$_7$ [Fig.~\ref{fig:2}(c)], it consists of Ac atoms and B$_2$H$_7$ units, where the bond length between B-B is 1.58 \AA\ at 100 GPa, and it is strongly covalently bonded with the ends coordinated to three H atoms and four H atoms. Interestingly, AcB$_2$H$_{14}$ contains B$_2$H$_8$ units, composed of two BH$_4$ linked by B-B covalent bonds, while three H$_2$ molecules in the same plane surround Ac atoms [Fig.~\ref{fig:2}(d)]. With increasing H contents, BH$_4$, BH$_5$, BH$_6$ and BH$_7$ appears in AcBH$_6$, AcB$_2$H$_{13}$, AcBH$_{16}$, AcBH$_7$ and AcBH$_8$, respectively [Figs.~\ref{fig:2}(e-i)]. In particular, AcBH$_7$ adopts high-symmetry $P\bar3m1$~[Fig.~\ref{fig:2}(h)], isostructral with LaBH$_7$~\cite{liang2021prediction}. The BH$_6$ units locate at the vertices and edges of the hexagonal lattice, and the ones on the edges are connected by H atoms. In addition to BH$_x$ units, the extra H atoms in Ac-B-H compounds exist in many forms, e.g., mono-atomic H (e.g., AcBH$_4$, AcB$_2$H$_7$, AcBH$_6$, AcB$_2$H$_8$, AcB$_2$H$_{14}$, AcB$_2$H$_{13}$, AcBH$_{16}$, AcBH$_7$, AcBH$_8$), H$_2$ (e.g., AcB$_2$H$_8$, AcB$_2$H$_{14}$, AcB$_2$H$_{13}$), and linear H$_3$ (e.g., AcB$_2$H$_{13}$).

 In order to further explore the fascinating properties of Ac-B-H compounds, we calculated their electronic structure and phonon properties. Almost all the predicted structures are metallic with several bands crossing the Fermi energy level, with the exception of AcBH$_6$ and AcB$_2$H$_7$, both of which are semiconductors (see Fig.~\ref{fig:3} and Figs. S8 and S9). The significant overlap of the partial electronic density of states (DOS) of the different atoms indicates a strong hybridization of Ac-H and B-H under pressure. Note that \Tc\ in BCS theory is closely related to the DOS of $E_f$. Additionally, the occupation of H at $E_f$ (DOS$_H$)  plays a crucial role as well. For the metallic structures, hydrogen atoms make a substantial contribution to the total DOS of $E_f$, e.g., for AcBH$_{16}$ (68\%), AcB$_2$H$_{13}$(64\%), AcBH$_7$(48\%), AcBH$_8$ (53\%), AcB$_2$H$_8$ (54\%) and AcB$_2$H$_{14}$ (58\%) (see Table S2). Considering these two criteria, five stable metallic of Ac-B-H compounds are screened as the potential superconductors: AcB$_2$H$_8$, AcB$_2$H$_{14}$, AcBH$_{16}$, AcBH$_7$ and AcBH$_8$.

Based on the metallic properties of Ac-B-H compounds, we calculated their phonon dispersions, projected phonon DOS (PHDOS), Eliashberg spectral function $\alpha^2F(\omega)$, and EPC integral $\lambda(\omega)$ at 200 GPa as shown in Fig.~\ref{fig:4} and Figs. S10-S12. The phonon dispersion supports their dynamical stability via the absence of any imaginary frequencies. Furthermore, the dynamical stability of LaBH$_7$ (at 85 GPa), AcBH$_7$ (at 70 GPa) and AcB$_2$H$_8$ (at 125 GPa) are confirmed by calculating PHDOS with larger $q$ meshes [Fig. S13]. The common feature of the phonons is that lowest frequency vibrational modes are contributed by the heavy atoms Ac, B and H dominate the mid-frequency modes, and the high-frequency modes are exclusively from H atoms. The superconducting properties were then evaluated using the Allen-Dynes-modified McMillan equation~\cite{allen1975transition} or by solving numerically the Eliashberg equations~\cite{eliashberg1960interactions}. We used the typical value of the Coulomb pseudopotential $\mu^\star$= 0.10. The results are  summarized in TABLE 1.

AcB$_2$H$_8$ exhibits good superconductivity with \Tc\ above 70 K at 200 GPa (Fig. S12). As the pressure decreases to 125 GPa, the superconductivity is almost unchanged with 79 K, which agrees well with the previous work~\cite{li2023pressure}. Other compounds possess the similar superconductivity with \Tc\ around 40 K at the corresponding predicted pressure.
AcBH$_{16}$ and AcB$_2$H$_{14}$ are dynamically unstable below 90 GPa and 180 GPa, respectively, thus their electronic properties are excluded from consideration at low pressure. Additionally, AcB$_2$H$_8$ has been reported previously, thus hereafter, our calculations and discussions will focus on AcBH$_8$ and AcBH$_7$, which could  remain dynamically stable at lower pressure.

 By further inspection, they have the close DOS at $E_f$ and H-DOS contribution at 200 GPa for AcBH$_8$ (0.030 eV$^{-1}$ per valence electron, 53\% from H-DOS) [Fig. S9(a)] and AcBH$_7$ (0.028 eV$^{-1}$ per valence electron, 48\% from H-DOS) [Fig.~\ref{fig:3}(b)]. Consequently, they exhibit very similar superconductivity. As the pressure decreases to 100 GPa, both the DOS at $E_f$ and H-contribution in AcBH$_8$ decrease accordingly (Table S2). Additionally, the phonons show obvious softening at high-frequencies under decompression (Fig. S10), which leads to an increase in the EPC parameter $\lambda$ from 0.64 to 0.71, while the average phonon frequency $\omega_\text{log}$ decreases from 1368 K to 1118 K, leading to the combined effect of a slight increase in \Tc\ from 43 K at 200 GPa to 46 K at 100 GPa.

 For AcBH$_7$, there are two bands crossing the Fermi energy level at 200 GPa with the largest contribution from H atoms to the total DOS [Figs.~\ref{fig:3} (a)-(b)]. When the pressure decreases to 70 GPa, the total DOS increases to 0.037 eV$^{-1}$ per valence electron, accompanied by an increase in H-DOS contribution [56\%, Fig.~\ref{fig:3}(b)], indicating the greater possibility to form the Cooper pairs that may contribute to higher \Tc. We further investigate the spin-orbit coupling (SOC) effect on the band structures of Ac-B-H compounds [Fig. 3(a) and Figs. S8-9], and found that they are nearly identical, indicating that the band structures are insensitive to SOC for actinium borohydrides.
 The calculated $\lambda$ and $\omega_\text{log}$ are 0.67 and 1309.04 K at 200 GPa, respectively, leading to a \Tc\ of 43 K. The low (0--250 cm$^{-1}$), mid (500--1000 cm$^{-1}$), and high frequencies (1000--2500 cm$^{-1}$) contribute 13\%, 28\%, and 59\% to the $\lambda$, respectively [Fig.~\ref{fig:4}(a)]. As the pressure is lowered to 70 GPa, $\lambda$ increases to 1.98, with a \Tc\ of 122 K. The enhanced superconductivity mainly originates from four softened modes labelled as "c", "d", "e", and "f". Among these, the contribution of low-frequency (0--250 cm$^{-1}$) induced by Ac atoms increases to 23\% to the total $\lambda$, due to the soft modes along L-H at site "c" [Fig.~\ref{fig:4}(b)]. More interestingly, the interaction between B and H becomes stronger, the mid-frequencies contribution to $\lambda$ due to soft phonon modes at "d", "e", and "f" (250--750 cm$^{-1}$) increases to 48\%. The detailed vibration modes of four soft modes at Fig.~\ref{fig:4}(b) are shown in Fig.~\ref{fig:4}(c-f), Specifically, the heavy Ac atoms contribute to the low-frequency modes at "c" site. For "e" and "f" sites, the softened vibrations modes are mainly originated from H atoms [Fig.~\ref{fig:4}(e-f)], while at "d" and "e" sites, the combined effect of B and H vibrations plays a key role in the phonon softening [see Fig.~\ref{fig:4}(d)].

 As compared with AcBH$_7$, isostructural LaBH$_7$ is calculated to remain dynamical stable at higher pressure (85 GPa), due to the smaller atomic mass of La atoms, which reduces the chemical pre-compression on BH$_6$ units. This is evidenced by comparing the lattice constant and B-H bond length in LaBH$_7$ and AcBH$_7$ (Fig. S14). The substitution of La atoms results in an increase in the lattice constants $a$ (= $b$), from 3.79~\AA~in AcBH$_7$ (70 GPa) to 3.91~\AA~(85 GPa) in LaBH$_7$, which is more obvious in the $c$ direction, from 5.21~\AA~to 5.75~\AA. Correspondingly, the length of the B-H bond in AcBH$_7$ is lightly shorter ($\sim$ 0.05\AA) than that in LaBH$_7$. Although the electron occupation at $E_f$ in LaBH$_7$ at 85 GPa (0.043 eV$^{-1}$ per valence electron) is larger than that of AcBH$_7$ (0.037 eV$^{-1}$ per valence electron) at 70 GPa. The contribution of H atoms is smaller with only 44\% [Fig. S15(a)], leading to the same H-DOS at $E_f$ with that of AcBH$_7$. Additionally, LaBH$_7$ has the same $\lambda$ (1.98), which is originated mainly from the contributions of B and H atoms at low and mid-frequencies (55\%) [Fig. S15(b)]. Consequently, they exhibit the same superconductivity with \Tc~of 120 K (Table S3).

 The effect of metal atoms on the superconductivity of this structure is further investigated by substituting the Ac atoms with its neighbouring metal atoms Th and Ce. Note that the space group $P\bar3m1$ evolves to $P6_3/mmc$ after full optimization. The superconducting parameters of CeBH$_7$ and ThBH$_7$ at their lowest dynamically stable pressures were summarized in Table S3. The onset pressure of dynamical stability for CeBH$_7$ (75 GPa) and ThBH$_7$ (50 GPa) is reduced compared to that of LaBH$_7$ and AcBH$_7$. It is reasonable considering the stronger chemical-compression from their larger atomic masses. However, their superconductivity is significantly weakened, only 18 and 51 K for CeBH$_7$ and ThBH$_7$, respectively. By comparing their electronic properties [Fig.~\ref{fig:5}(a)-(b)], we find that both ThBH$_7$ and CeBH$_7$ have larger total DOS at $E_f$, however, the contribution of H atoms to the DOS is much smaller in the metal atoms Ce and Th. This is obviously not favourable to obtain good superconductivity. Thus their phonons display very similar features, which are also divided into three parts, but with much smaller $\lambda$, 0.61 and 1.04 for CeBH$_7$ and ThBH$_7$, respectively [Fig.~\ref{fig:5}(c)-(d) and Fig. S16]. These results further demonstrate the key role of H-dominated DOS at $E_f$ in attaining high-\Tc\ superconductors.

\section{CONCLUSIONS}
In summary, the crystal structure and superconductivity of the Ac-B-H system at 100~GPa and 200~GPa are explored by a combination of structure prediction methods and first principles calculations. We uncover nine thermodynamically stable structures, namely: AcBH, AcBH$_4$, AcBH$_6$, AcBH$_{16}$, AcB$_2$H$_7$, AcB$_2$H$_{13}$, AcBH$_7$, AcBH$_8$, AcB$_2$H$_8$ and AcB$_2$H$_{14}$. Seven of them exhibit metallic nature. Electron-phonon coupling calculations show that AcBH$_7$ is a potential superconductor with \Tc\ of 43 K at 200~GPa. Interestingly, it remains dynamically stable down to 70~GPa with an increased \Tc\ of 122 K. The enhanced superconductivity is mainly attributed to the softened phonon vibrations originating from Ac atoms at low frequency, as well as B-H interactions in BH$_6$ units at mid-frequency. Our work is expected to stimulate more research on ternary superconducting hydrides in the hope to uncover more compounds with high critical temperature and stability at low pressure.

\section{Acknowledgments}

The authors acknowledge funding from the NSFC under Grants No. 12074154, No. 12174160, No. 12074138, No. 52288102, No. 52090024, and No. 11722433. Y.L. acknowledges the funding from the Six Talent Peaks Project and 333 High-level Talents Project of Jiangsu Province.  H.L. acknowledges the funding from the Strategic Priority Research Program of Chinese Academy of Sciences (Grant No. XDB33000000). A.P.D. is grateful for financial support from the National Science Centre (Poland) through Project No. 2022/47/B/ST3/00622. All the calculations were performed at the High Performance Computing Center of the School of Physics and Electronic Engineering of Jiangsu Normal University.
\bibliography{AcBH}

\begin{thebibliography}{68}%
\makeatletter
\providecommand \@ifxundefined [1]{%
 \@ifx{#1\undefined}
}%
\providecommand \@ifnum [1]{%
 \ifnum #1\expandafter \@firstoftwo
 \else \expandafter \@secondoftwo
 \fi
}%
\providecommand \@ifx [1]{%
 \ifx #1\expandafter \@firstoftwo
 \else \expandafter \@secondoftwo
 \fi
}%
\providecommand \natexlab [1]{#1}%
\providecommand \enquote  [1]{``#1''}%
\providecommand \bibnamefont  [1]{#1}%
\providecommand \bibfnamefont [1]{#1}%
\providecommand \citenamefont [1]{#1}%
\providecommand \href@noop [0]{\@secondoftwo}%
\providecommand \href [0]{\begingroup \@sanitize@url \@href}%
\providecommand \@href[1]{\@@startlink{#1}\@@href}%
\providecommand \@@href[1]{\endgroup#1\@@endlink}%
\providecommand \@sanitize@url [0]{\catcode `\\12\catcode `\$12\catcode
  `\&12\catcode `\#12\catcode `\^12\catcode `\_12\catcode `\%12\relax}%
\providecommand \@@startlink[1]{}%
\providecommand \@@endlink[0]{}%
\providecommand \url  [0]{\begingroup\@sanitize@url \@url }%
\providecommand \@url [1]{\endgroup\@href {#1}{\urlprefix }}%
\providecommand \urlprefix  [0]{URL }%
\providecommand \Eprint [0]{\href }%
\providecommand \doibase [0]{http://dx.doi.org/}%
\providecommand \selectlanguage [0]{\@gobble}%
\providecommand \bibinfo  [0]{\@secondoftwo}%
\providecommand \bibfield  [0]{\@secondoftwo}%
\providecommand \translation [1]{[#1]}%
\providecommand \BibitemOpen [0]{}%
\providecommand \bibitemStop [0]{}%
\providecommand \bibitemNoStop [0]{.\EOS\space}%
\providecommand \EOS [0]{\spacefactor3000\relax}%
\providecommand \BibitemShut  [1]{\csname bibitem#1\endcsname}%
\let\auto@bib@innerbib\@empty
\bibitem [{\citenamefont {Wigner}\ and\ \citenamefont
  {Huntington}(1935)}]{wigner1935possibility}%
  \BibitemOpen
  \bibfield  {author} {\bibinfo {author} {\bibfnamefont {E.}~\bibnamefont
  {Wigner}}\ and\ \bibinfo {author} {\bibfnamefont {H.~{\'a}.}\ \bibnamefont
  {Huntington}},\ }\href@noop {} {\bibfield  {journal} {\bibinfo  {journal} {J.
  Chem. Phys.}\ }\textbf {\bibinfo {volume} {3}},\ \bibinfo {pages} {764}
  (\bibinfo {year} {1935})}\BibitemShut {NoStop}%
\bibitem [{\citenamefont {Ashcroft}(1968)}]{ashcroft1968metallic}%
  \BibitemOpen
  \bibfield  {author} {\bibinfo {author} {\bibfnamefont {N.~W.}\ \bibnamefont
  {Ashcroft}},\ }\href@noop {} {\bibfield  {journal} {\bibinfo  {journal}
  {Phys. Rev. Lett.}\ }\textbf {\bibinfo {volume} {21}},\ \bibinfo {pages}
  {1748} (\bibinfo {year} {1968})}\BibitemShut {NoStop}%
\bibitem [{\citenamefont {Azadi}\ \emph {et~al.}(2014)\citenamefont {Azadi},
  \citenamefont {Monserrat}, \citenamefont {Foulkes},\ and\ \citenamefont
  {Needs}}]{azadi2014dissociation}%
  \BibitemOpen
  \bibfield  {author} {\bibinfo {author} {\bibfnamefont {S.}~\bibnamefont
  {Azadi}}, \bibinfo {author} {\bibfnamefont {B.}~\bibnamefont {Monserrat}},
  \bibinfo {author} {\bibfnamefont {W.}~\bibnamefont {Foulkes}}, \ and\
  \bibinfo {author} {\bibfnamefont {R.}~\bibnamefont {Needs}},\ }\href@noop {}
  {\bibfield  {journal} {\bibinfo  {journal} {Phys. Rev. Lett.}\ }\textbf
  {\bibinfo {volume} {112}},\ \bibinfo {pages} {165501} (\bibinfo {year}
  {2014})}\BibitemShut {NoStop}%
\bibitem [{\citenamefont {McMinis}\ \emph {et~al.}(2015)\citenamefont
  {McMinis}, \citenamefont {Clay~III}, \citenamefont {Lee},\ and\ \citenamefont
  {Morales}}]{mcminis2015molecular}%
  \BibitemOpen
  \bibfield  {author} {\bibinfo {author} {\bibfnamefont {J.}~\bibnamefont
  {McMinis}}, \bibinfo {author} {\bibfnamefont {R.~C.}\ \bibnamefont
  {Clay~III}}, \bibinfo {author} {\bibfnamefont {D.}~\bibnamefont {Lee}}, \
  and\ \bibinfo {author} {\bibfnamefont {M.~A.}\ \bibnamefont {Morales}},\
  }\href@noop {} {\bibfield  {journal} {\bibinfo  {journal} {Phys. Rev. Lett.}\
  }\textbf {\bibinfo {volume} {114}},\ \bibinfo {pages} {105305} (\bibinfo
  {year} {2015})}\BibitemShut {NoStop}%
\bibitem [{\citenamefont {Ashcroft}(2004)}]{ashcroft2004hydrogen}%
  \BibitemOpen
  \bibfield  {author} {\bibinfo {author} {\bibfnamefont {N.}~\bibnamefont
  {Ashcroft}},\ }\href@noop {} {\bibfield  {journal} {\bibinfo  {journal}
  {Phys. Rev. Lett.}\ }\textbf {\bibinfo {volume} {92}},\ \bibinfo {pages}
  {187002} (\bibinfo {year} {2004})}\BibitemShut {NoStop}%
\bibitem [{\citenamefont {Li}\ \emph {et~al.}(2014)\citenamefont {Li},
  \citenamefont {Hao}, \citenamefont {Liu}, \citenamefont {Li},\ and\
  \citenamefont {Ma}}]{li2014metallization}%
  \BibitemOpen
  \bibfield  {author} {\bibinfo {author} {\bibfnamefont {Y.}~\bibnamefont
  {Li}}, \bibinfo {author} {\bibfnamefont {J.}~\bibnamefont {Hao}}, \bibinfo
  {author} {\bibfnamefont {H.}~\bibnamefont {Liu}}, \bibinfo {author}
  {\bibfnamefont {Y.}~\bibnamefont {Li}}, \ and\ \bibinfo {author}
  {\bibfnamefont {Y.}~\bibnamefont {Ma}},\ }\href@noop {} {\bibfield  {journal}
  {\bibinfo  {journal} {J. Chem. Phys.}\ }\textbf {\bibinfo {volume} {140}},\
  \bibinfo {pages} {174712} (\bibinfo {year} {2014})}\BibitemShut {NoStop}%
\bibitem [{\citenamefont {Duan}\ \emph {et~al.}(2015)\citenamefont {Duan},
  \citenamefont {Huang}, \citenamefont {Tian}, \citenamefont {Li},
  \citenamefont {Yu}, \citenamefont {Liu}, \citenamefont {Ma}, \citenamefont
  {Liu},\ and\ \citenamefont {Cui}}]{duan2015pressure}%
  \BibitemOpen
  \bibfield  {author} {\bibinfo {author} {\bibfnamefont {D.}~\bibnamefont
  {Duan}}, \bibinfo {author} {\bibfnamefont {X.}~\bibnamefont {Huang}},
  \bibinfo {author} {\bibfnamefont {F.}~\bibnamefont {Tian}}, \bibinfo {author}
  {\bibfnamefont {D.}~\bibnamefont {Li}}, \bibinfo {author} {\bibfnamefont
  {H.}~\bibnamefont {Yu}}, \bibinfo {author} {\bibfnamefont {Y.}~\bibnamefont
  {Liu}}, \bibinfo {author} {\bibfnamefont {Y.}~\bibnamefont {Ma}}, \bibinfo
  {author} {\bibfnamefont {B.}~\bibnamefont {Liu}}, \ and\ \bibinfo {author}
  {\bibfnamefont {T.}~\bibnamefont {Cui}},\ }\href@noop {} {\bibfield
  {journal} {\bibinfo  {journal} {Phys. Rev. B}\ }\textbf {\bibinfo {volume}
  {91}},\ \bibinfo {pages} {180502} (\bibinfo {year} {2015})}\BibitemShut
  {NoStop}%
\bibitem [{\citenamefont {Duan}\ \emph {et~al.}()\citenamefont {Duan},
  \citenamefont {Liu}, \citenamefont {Tian}, \citenamefont {Li}, \citenamefont
  {Huang}, \citenamefont {Zhao}, \citenamefont {Yu}, \citenamefont {Liu},
  \citenamefont {Tian},\ and\ \citenamefont {Cui}}]{duan2014pressure}%
  \BibitemOpen
  \bibfield  {author} {\bibinfo {author} {\bibfnamefont {D.}~\bibnamefont
  {Duan}}, \bibinfo {author} {\bibfnamefont {Y.}~\bibnamefont {Liu}}, \bibinfo
  {author} {\bibfnamefont {F.}~\bibnamefont {Tian}}, \bibinfo {author}
  {\bibfnamefont {D.}~\bibnamefont {Li}}, \bibinfo {author} {\bibfnamefont
  {X.}~\bibnamefont {Huang}}, \bibinfo {author} {\bibfnamefont
  {Z.}~\bibnamefont {Zhao}}, \bibinfo {author} {\bibfnamefont {H.}~\bibnamefont
  {Yu}}, \bibinfo {author} {\bibfnamefont {B.}~\bibnamefont {Liu}}, \bibinfo
  {author} {\bibfnamefont {W.}~\bibnamefont {Tian}}, \ and\ \bibinfo {author}
  {\bibfnamefont {T.}~\bibnamefont {Cui}},\ }\href@noop {} {\bibfield
  {journal} {\bibinfo  {journal} {Sci. Rep.}\ }\textbf {\bibinfo {volume}
  {4}}}\BibitemShut {NoStop}%
\bibitem [{\citenamefont {Drozdov}\ \emph {et~al.}(2015)\citenamefont
  {Drozdov}, \citenamefont {Eremets}, \citenamefont {Troyan}, \citenamefont
  {Ksenofontov},\ and\ \citenamefont {Shylin}}]{drozdov2015conventional}%
  \BibitemOpen
  \bibfield  {author} {\bibinfo {author} {\bibfnamefont {A.}~\bibnamefont
  {Drozdov}}, \bibinfo {author} {\bibfnamefont {M.}~\bibnamefont {Eremets}},
  \bibinfo {author} {\bibfnamefont {I.}~\bibnamefont {Troyan}}, \bibinfo
  {author} {\bibfnamefont {V.}~\bibnamefont {Ksenofontov}}, \ and\ \bibinfo
  {author} {\bibfnamefont {S.~I.}\ \bibnamefont {Shylin}},\ }\href@noop {}
  {\bibfield  {journal} {\bibinfo  {journal} {Nature}\ }\textbf {\bibinfo
  {volume} {525}},\ \bibinfo {pages} {73} (\bibinfo {year} {2015})}\BibitemShut
  {NoStop}%
\bibitem [{\citenamefont {Liu}\ \emph {et~al.}(2017)\citenamefont {Liu},
  \citenamefont {Naumov}, \citenamefont {Hoffmann}, \citenamefont {Ashcroft},\
  and\ \citenamefont {Hemley}}]{liu2017potential}%
  \BibitemOpen
  \bibfield  {author} {\bibinfo {author} {\bibfnamefont {H.}~\bibnamefont
  {Liu}}, \bibinfo {author} {\bibfnamefont {I.~I.}\ \bibnamefont {Naumov}},
  \bibinfo {author} {\bibfnamefont {R.}~\bibnamefont {Hoffmann}}, \bibinfo
  {author} {\bibfnamefont {N.}~\bibnamefont {Ashcroft}}, \ and\ \bibinfo
  {author} {\bibfnamefont {R.~J.}\ \bibnamefont {Hemley}},\ }\href@noop {}
  {\bibfield  {journal} {\bibinfo  {journal} {Proc. Natl. Acad. Sci.}\ }\textbf
  {\bibinfo {volume} {114}},\ \bibinfo {pages} {6990} (\bibinfo {year}
  {2017})}\BibitemShut {NoStop}%
\bibitem [{\citenamefont {Drozdov}\ \emph {et~al.}(2019)\citenamefont
  {Drozdov}, \citenamefont {Kong}, \citenamefont {Minkov}, \citenamefont
  {Besedin}, \citenamefont {Kuzovnikov}, \citenamefont {Mozaffari},
  \citenamefont {Balicas}, \citenamefont {Balakirev}, \citenamefont {Graf},
  \citenamefont {Prakapenka} \emph {et~al.}}]{drozdov2019superconductivity}%
  \BibitemOpen
  \bibfield  {author} {\bibinfo {author} {\bibfnamefont {A.}~\bibnamefont
  {Drozdov}}, \bibinfo {author} {\bibfnamefont {P.}~\bibnamefont {Kong}},
  \bibinfo {author} {\bibfnamefont {V.}~\bibnamefont {Minkov}}, \bibinfo
  {author} {\bibfnamefont {S.}~\bibnamefont {Besedin}}, \bibinfo {author}
  {\bibfnamefont {M.}~\bibnamefont {Kuzovnikov}}, \bibinfo {author}
  {\bibfnamefont {S.}~\bibnamefont {Mozaffari}}, \bibinfo {author}
  {\bibfnamefont {L.}~\bibnamefont {Balicas}}, \bibinfo {author} {\bibfnamefont
  {F.}~\bibnamefont {Balakirev}}, \bibinfo {author} {\bibfnamefont
  {D.}~\bibnamefont {Graf}}, \bibinfo {author} {\bibfnamefont {V.}~\bibnamefont
  {Prakapenka}},  \emph {et~al.},\ }\href@noop {} {\bibfield  {journal}
  {\bibinfo  {journal} {Nature}\ }\textbf {\bibinfo {volume} {569}},\ \bibinfo
  {pages} {528} (\bibinfo {year} {2019})}\BibitemShut {NoStop}%
\bibitem [{\citenamefont {Peng}\ \emph {et~al.}(2017)\citenamefont {Peng},
  \citenamefont {Sun}, \citenamefont {Pickard}, \citenamefont {Needs},
  \citenamefont {Wu},\ and\ \citenamefont {Ma}}]{peng2017hydrogen}%
  \BibitemOpen
  \bibfield  {author} {\bibinfo {author} {\bibfnamefont {F.}~\bibnamefont
  {Peng}}, \bibinfo {author} {\bibfnamefont {Y.}~\bibnamefont {Sun}}, \bibinfo
  {author} {\bibfnamefont {C.~J.}\ \bibnamefont {Pickard}}, \bibinfo {author}
  {\bibfnamefont {R.~J.}\ \bibnamefont {Needs}}, \bibinfo {author}
  {\bibfnamefont {Q.}~\bibnamefont {Wu}}, \ and\ \bibinfo {author}
  {\bibfnamefont {Y.}~\bibnamefont {Ma}},\ }\href@noop {} {\bibfield  {journal}
  {\bibinfo  {journal} {Phys. Rev. Lett.}\ }\textbf {\bibinfo {volume} {119}},\
  \bibinfo {pages} {107001} (\bibinfo {year} {2017})}\BibitemShut {NoStop}%
\bibitem [{\citenamefont {Somayazulu}\ \emph {et~al.}(2019)\citenamefont
  {Somayazulu}, \citenamefont {Ahart}, \citenamefont {Mishra}, \citenamefont
  {Geballe}, \citenamefont {Baldini}, \citenamefont {Meng}, \citenamefont
  {Struzhkin},\ and\ \citenamefont {Hemley}}]{somayazulu2019evidence}%
  \BibitemOpen
  \bibfield  {author} {\bibinfo {author} {\bibfnamefont {M.}~\bibnamefont
  {Somayazulu}}, \bibinfo {author} {\bibfnamefont {M.}~\bibnamefont {Ahart}},
  \bibinfo {author} {\bibfnamefont {A.~K.}\ \bibnamefont {Mishra}}, \bibinfo
  {author} {\bibfnamefont {Z.~M.}\ \bibnamefont {Geballe}}, \bibinfo {author}
  {\bibfnamefont {M.}~\bibnamefont {Baldini}}, \bibinfo {author} {\bibfnamefont
  {Y.}~\bibnamefont {Meng}}, \bibinfo {author} {\bibfnamefont {V.~V.}\
  \bibnamefont {Struzhkin}}, \ and\ \bibinfo {author} {\bibfnamefont {R.~J.}\
  \bibnamefont {Hemley}},\ }\href@noop {} {\bibfield  {journal} {\bibinfo
  {journal} {Phys. Rev. Lett.}\ }\textbf {\bibinfo {volume} {122}},\ \bibinfo
  {pages} {027001} (\bibinfo {year} {2019})}\BibitemShut {NoStop}%
\bibitem [{\citenamefont {Li}\ \emph {et~al.}(2015)\citenamefont {Li},
  \citenamefont {Hao}, \citenamefont {Liu}, \citenamefont {Tse}, \citenamefont
  {Wang},\ and\ \citenamefont {Ma}}]{li2015pressure}%
  \BibitemOpen
  \bibfield  {author} {\bibinfo {author} {\bibfnamefont {Y.}~\bibnamefont
  {Li}}, \bibinfo {author} {\bibfnamefont {J.}~\bibnamefont {Hao}}, \bibinfo
  {author} {\bibfnamefont {H.}~\bibnamefont {Liu}}, \bibinfo {author}
  {\bibfnamefont {J.~S.}\ \bibnamefont {Tse}}, \bibinfo {author} {\bibfnamefont
  {Y.}~\bibnamefont {Wang}}, \ and\ \bibinfo {author} {\bibfnamefont
  {Y.}~\bibnamefont {Ma}},\ }\href@noop {} {\bibfield  {journal} {\bibinfo
  {journal} {Sci. Rep.}\ }\textbf {\bibinfo {volume} {5}},\ \bibinfo {pages}
  {9948} (\bibinfo {year} {2015})}\BibitemShut {NoStop}%
\bibitem [{\citenamefont {Kong}\ \emph {et~al.}(2021)\citenamefont {Kong},
  \citenamefont {Minkov}, \citenamefont {Kuzovnikov}, \citenamefont {Drozdov},
  \citenamefont {Besedin}, \citenamefont {Mozaffari}, \citenamefont {Balicas},
  \citenamefont {Balakirev}, \citenamefont {Prakapenka}, \citenamefont
  {Chariton} \emph {et~al.}}]{kong2021superconductivity}%
  \BibitemOpen
  \bibfield  {author} {\bibinfo {author} {\bibfnamefont {P.}~\bibnamefont
  {Kong}}, \bibinfo {author} {\bibfnamefont {V.~S.}\ \bibnamefont {Minkov}},
  \bibinfo {author} {\bibfnamefont {M.~A.}\ \bibnamefont {Kuzovnikov}},
  \bibinfo {author} {\bibfnamefont {A.~P.}\ \bibnamefont {Drozdov}}, \bibinfo
  {author} {\bibfnamefont {S.~P.}\ \bibnamefont {Besedin}}, \bibinfo {author}
  {\bibfnamefont {S.}~\bibnamefont {Mozaffari}}, \bibinfo {author}
  {\bibfnamefont {L.}~\bibnamefont {Balicas}}, \bibinfo {author} {\bibfnamefont
  {F.~F.}\ \bibnamefont {Balakirev}}, \bibinfo {author} {\bibfnamefont {V.~B.}\
  \bibnamefont {Prakapenka}}, \bibinfo {author} {\bibfnamefont
  {S.}~\bibnamefont {Chariton}},  \emph {et~al.},\ }\href@noop {} {\bibfield
  {journal} {\bibinfo  {journal} {Nat. Commun.}\ }\textbf {\bibinfo {volume}
  {12}},\ \bibinfo {pages} {5075} (\bibinfo {year} {2021})}\BibitemShut
  {NoStop}%
\bibitem [{\citenamefont {Troyan}\ \emph {et~al.}(2021)\citenamefont {Troyan},
  \citenamefont {Semenok}, \citenamefont {Kvashnin}, \citenamefont {Sadakov},
  \citenamefont {Sobolevskiy}, \citenamefont {Pudalov}, \citenamefont
  {Ivanova}, \citenamefont {Prakapenka}, \citenamefont {Greenberg},
  \citenamefont {Gavriliuk} \emph {et~al.}}]{troyan2021anomalous}%
  \BibitemOpen
  \bibfield  {author} {\bibinfo {author} {\bibfnamefont {I.~A.}\ \bibnamefont
  {Troyan}}, \bibinfo {author} {\bibfnamefont {D.~V.}\ \bibnamefont {Semenok}},
  \bibinfo {author} {\bibfnamefont {A.~G.}\ \bibnamefont {Kvashnin}}, \bibinfo
  {author} {\bibfnamefont {A.~V.}\ \bibnamefont {Sadakov}}, \bibinfo {author}
  {\bibfnamefont {O.~A.}\ \bibnamefont {Sobolevskiy}}, \bibinfo {author}
  {\bibfnamefont {V.~M.}\ \bibnamefont {Pudalov}}, \bibinfo {author}
  {\bibfnamefont {A.~G.}\ \bibnamefont {Ivanova}}, \bibinfo {author}
  {\bibfnamefont {V.~B.}\ \bibnamefont {Prakapenka}}, \bibinfo {author}
  {\bibfnamefont {E.}~\bibnamefont {Greenberg}}, \bibinfo {author}
  {\bibfnamefont {A.~G.}\ \bibnamefont {Gavriliuk}},  \emph {et~al.},\
  }\href@noop {} {\bibfield  {journal} {\bibinfo  {journal} {Adv. Mater.}\
  }\textbf {\bibinfo {volume} {33}},\ \bibinfo {pages} {2006832} (\bibinfo
  {year} {2021})}\BibitemShut {NoStop}%
\bibitem [{\citenamefont {Snider}\ \emph {et~al.}(2021)\citenamefont {Snider},
  \citenamefont {Dasenbrock-Gammon}, \citenamefont {McBride}, \citenamefont
  {Wang}, \citenamefont {Meyers}, \citenamefont {Lawler}, \citenamefont
  {Zurek}, \citenamefont {Salamat},\ and\ \citenamefont
  {Dias}}]{snider2021synthesis}%
  \BibitemOpen
  \bibfield  {author} {\bibinfo {author} {\bibfnamefont {E.}~\bibnamefont
  {Snider}}, \bibinfo {author} {\bibfnamefont {N.}~\bibnamefont
  {Dasenbrock-Gammon}}, \bibinfo {author} {\bibfnamefont {R.}~\bibnamefont
  {McBride}}, \bibinfo {author} {\bibfnamefont {X.}~\bibnamefont {Wang}},
  \bibinfo {author} {\bibfnamefont {N.}~\bibnamefont {Meyers}}, \bibinfo
  {author} {\bibfnamefont {K.~V.}\ \bibnamefont {Lawler}}, \bibinfo {author}
  {\bibfnamefont {E.}~\bibnamefont {Zurek}}, \bibinfo {author} {\bibfnamefont
  {A.}~\bibnamefont {Salamat}}, \ and\ \bibinfo {author} {\bibfnamefont
  {R.~P.}\ \bibnamefont {Dias}},\ }\href@noop {} {\bibfield  {journal}
  {\bibinfo  {journal} {Phys. Rev. Lett.}\ }\textbf {\bibinfo {volume} {126}},\
  \bibinfo {pages} {117003} (\bibinfo {year} {2021})}\BibitemShut {NoStop}%
\bibitem [{\citenamefont {Wang}\ \emph
  {et~al.}(2012{\natexlab{a}})\citenamefont {Wang}, \citenamefont {Tse},
  \citenamefont {Tanaka}, \citenamefont {Iitaka},\ and\ \citenamefont
  {Ma}}]{wang2012superconductive}%
  \BibitemOpen
  \bibfield  {author} {\bibinfo {author} {\bibfnamefont {H.}~\bibnamefont
  {Wang}}, \bibinfo {author} {\bibfnamefont {J.~S.}\ \bibnamefont {Tse}},
  \bibinfo {author} {\bibfnamefont {K.}~\bibnamefont {Tanaka}}, \bibinfo
  {author} {\bibfnamefont {T.}~\bibnamefont {Iitaka}}, \ and\ \bibinfo {author}
  {\bibfnamefont {Y.}~\bibnamefont {Ma}},\ }\href@noop {} {\bibfield  {journal}
  {\bibinfo  {journal} {Proc. Natl. Acad. Sci.}\ }\textbf {\bibinfo {volume}
  {109}},\ \bibinfo {pages} {6463} (\bibinfo {year}
  {2012}{\natexlab{a}})}\BibitemShut {NoStop}%
\bibitem [{\citenamefont {Ma}\ \emph {et~al.}(2022)\citenamefont {Ma},
  \citenamefont {Wang}, \citenamefont {Xie}, \citenamefont {Yang},
  \citenamefont {Wang}, \citenamefont {Zhou}, \citenamefont {Liu},
  \citenamefont {Yu}, \citenamefont {Zhao}, \citenamefont {Wang} \emph
  {et~al.}}]{ma2022high}%
  \BibitemOpen
  \bibfield  {author} {\bibinfo {author} {\bibfnamefont {L.}~\bibnamefont
  {Ma}}, \bibinfo {author} {\bibfnamefont {K.}~\bibnamefont {Wang}}, \bibinfo
  {author} {\bibfnamefont {Y.}~\bibnamefont {Xie}}, \bibinfo {author}
  {\bibfnamefont {X.}~\bibnamefont {Yang}}, \bibinfo {author} {\bibfnamefont
  {Y.}~\bibnamefont {Wang}}, \bibinfo {author} {\bibfnamefont {M.}~\bibnamefont
  {Zhou}}, \bibinfo {author} {\bibfnamefont {H.}~\bibnamefont {Liu}}, \bibinfo
  {author} {\bibfnamefont {X.}~\bibnamefont {Yu}}, \bibinfo {author}
  {\bibfnamefont {Y.}~\bibnamefont {Zhao}}, \bibinfo {author} {\bibfnamefont
  {H.}~\bibnamefont {Wang}},  \emph {et~al.},\ }\href@noop {} {\bibfield
  {journal} {\bibinfo  {journal} {Phys. Rev. Lett.}\ }\textbf {\bibinfo
  {volume} {128}},\ \bibinfo {pages} {167001} (\bibinfo {year}
  {2022})}\BibitemShut {NoStop}%
\bibitem [{\citenamefont {Li}\ \emph {et~al.}(2022{\natexlab{a}})\citenamefont
  {Li}, \citenamefont {He}, \citenamefont {Zhang}, \citenamefont {Wang},
  \citenamefont {Zhang}, \citenamefont {Jia}, \citenamefont {Feng},
  \citenamefont {Lu}, \citenamefont {Zhao}, \citenamefont {Zhang} \emph
  {et~al.}}]{li2022superconductivity1}%
  \BibitemOpen
  \bibfield  {author} {\bibinfo {author} {\bibfnamefont {Z.}~\bibnamefont
  {Li}}, \bibinfo {author} {\bibfnamefont {X.}~\bibnamefont {He}}, \bibinfo
  {author} {\bibfnamefont {C.}~\bibnamefont {Zhang}}, \bibinfo {author}
  {\bibfnamefont {X.}~\bibnamefont {Wang}}, \bibinfo {author} {\bibfnamefont
  {S.}~\bibnamefont {Zhang}}, \bibinfo {author} {\bibfnamefont
  {Y.}~\bibnamefont {Jia}}, \bibinfo {author} {\bibfnamefont {S.}~\bibnamefont
  {Feng}}, \bibinfo {author} {\bibfnamefont {K.}~\bibnamefont {Lu}}, \bibinfo
  {author} {\bibfnamefont {J.}~\bibnamefont {Zhao}}, \bibinfo {author}
  {\bibfnamefont {J.}~\bibnamefont {Zhang}},  \emph {et~al.},\ }\href@noop {}
  {\bibfield  {journal} {\bibinfo  {journal} {Nat. Commun.}\ }\textbf {\bibinfo
  {volume} {13}},\ \bibinfo {pages} {2863} (\bibinfo {year}
  {2022}{\natexlab{a}})}\BibitemShut {NoStop}%
\bibitem [{\citenamefont {Zurek}\ and\ \citenamefont
  {Bi}(2019)}]{zurek2019high}%
  \BibitemOpen
  \bibfield  {author} {\bibinfo {author} {\bibfnamefont {E.}~\bibnamefont
  {Zurek}}\ and\ \bibinfo {author} {\bibfnamefont {T.}~\bibnamefont {Bi}},\
  }\href@noop {} {\bibfield  {journal} {\bibinfo  {journal} {J. Chem. Phys.}\
  }\textbf {\bibinfo {volume} {150}},\ \bibinfo {pages} {050901} (\bibinfo
  {year} {2019})}\BibitemShut {NoStop}%
\bibitem [{\citenamefont {Flores-Livas}\ \emph {et~al.}(2020)\citenamefont
  {Flores-Livas}, \citenamefont {Boeri}, \citenamefont {Sanna}, \citenamefont
  {Profeta}, \citenamefont {Arita},\ and\ \citenamefont
  {Eremets}}]{flores2020perspective}%
  \BibitemOpen
  \bibfield  {author} {\bibinfo {author} {\bibfnamefont {J.~A.}\ \bibnamefont
  {Flores-Livas}}, \bibinfo {author} {\bibfnamefont {L.}~\bibnamefont {Boeri}},
  \bibinfo {author} {\bibfnamefont {A.}~\bibnamefont {Sanna}}, \bibinfo
  {author} {\bibfnamefont {G.}~\bibnamefont {Profeta}}, \bibinfo {author}
  {\bibfnamefont {R.}~\bibnamefont {Arita}}, \ and\ \bibinfo {author}
  {\bibfnamefont {M.}~\bibnamefont {Eremets}},\ }\href@noop {} {\bibfield
  {journal} {\bibinfo  {journal} {Phys. Rep.}\ }\textbf {\bibinfo {volume}
  {856}},\ \bibinfo {pages} {1} (\bibinfo {year} {2020})}\BibitemShut {NoStop}%
\bibitem [{\citenamefont {Semenok}\ \emph {et~al.}(2021)\citenamefont
  {Semenok}, \citenamefont {Troyan}, \citenamefont {Ivanova}, \citenamefont
  {Kvashnin}, \citenamefont {Kruglov}, \citenamefont {Hanfland}, \citenamefont
  {Sadakov}, \citenamefont {Sobolevskiy}, \citenamefont {Pervakov},
  \citenamefont {Lyubutin} \emph {et~al.}}]{semenok2021superconductivity}%
  \BibitemOpen
  \bibfield  {author} {\bibinfo {author} {\bibfnamefont {D.~V.}\ \bibnamefont
  {Semenok}}, \bibinfo {author} {\bibfnamefont {I.~A.}\ \bibnamefont {Troyan}},
  \bibinfo {author} {\bibfnamefont {A.~G.}\ \bibnamefont {Ivanova}}, \bibinfo
  {author} {\bibfnamefont {A.~G.}\ \bibnamefont {Kvashnin}}, \bibinfo {author}
  {\bibfnamefont {I.~A.}\ \bibnamefont {Kruglov}}, \bibinfo {author}
  {\bibfnamefont {M.}~\bibnamefont {Hanfland}}, \bibinfo {author}
  {\bibfnamefont {A.~V.}\ \bibnamefont {Sadakov}}, \bibinfo {author}
  {\bibfnamefont {O.~A.}\ \bibnamefont {Sobolevskiy}}, \bibinfo {author}
  {\bibfnamefont {K.~S.}\ \bibnamefont {Pervakov}}, \bibinfo {author}
  {\bibfnamefont {I.~S.}\ \bibnamefont {Lyubutin}},  \emph {et~al.},\
  }\href@noop {} {\bibfield  {journal} {\bibinfo  {journal} {Mater. Today}\
  }\textbf {\bibinfo {volume} {48}},\ \bibinfo {pages} {18} (\bibinfo {year}
  {2021})}\BibitemShut {NoStop}%
\bibitem [{\citenamefont {Chen}\ \emph {et~al.}(2022)\citenamefont {Chen},
  \citenamefont {Huang}, \citenamefont {Semenok}, \citenamefont {Chen},
  \citenamefont {Zhou}, \citenamefont {Zhang}, \citenamefont {Oganov},\ and\
  \citenamefont {Cui}}]{chen2022enhancement}%
  \BibitemOpen
  \bibfield  {author} {\bibinfo {author} {\bibfnamefont {W.}~\bibnamefont
  {Chen}}, \bibinfo {author} {\bibfnamefont {X.}~\bibnamefont {Huang}},
  \bibinfo {author} {\bibfnamefont {D.}~\bibnamefont {Semenok}}, \bibinfo
  {author} {\bibfnamefont {S.}~\bibnamefont {Chen}}, \bibinfo {author}
  {\bibfnamefont {D.}~\bibnamefont {Zhou}}, \bibinfo {author} {\bibfnamefont
  {K.}~\bibnamefont {Zhang}}, \bibinfo {author} {\bibfnamefont
  {A.}~\bibnamefont {Oganov}}, \ and\ \bibinfo {author} {\bibfnamefont
  {T.}~\bibnamefont {Cui}},\ }\href@noop {} {\  (\bibinfo {year}
  {2022})}\BibitemShut {NoStop}%
\bibitem [{\citenamefont {Bi}\ \emph {et~al.}(2022)\citenamefont {Bi},
  \citenamefont {Nakamoto}, \citenamefont {Shimizu}, \citenamefont {Zhou},
  \citenamefont {Wang}, \citenamefont {Liu},\ and\ \citenamefont
  {Ma}}]{bi2022efficient}%
  \BibitemOpen
  \bibfield  {author} {\bibinfo {author} {\bibfnamefont {J.}~\bibnamefont
  {Bi}}, \bibinfo {author} {\bibfnamefont {Y.}~\bibnamefont {Nakamoto}},
  \bibinfo {author} {\bibfnamefont {K.}~\bibnamefont {Shimizu}}, \bibinfo
  {author} {\bibfnamefont {M.}~\bibnamefont {Zhou}}, \bibinfo {author}
  {\bibfnamefont {H.}~\bibnamefont {Wang}}, \bibinfo {author} {\bibfnamefont
  {G.}~\bibnamefont {Liu}}, \ and\ \bibinfo {author} {\bibfnamefont
  {Y.}~\bibnamefont {Ma}},\ }\href@noop {} {\bibfield  {journal} {\bibinfo
  {journal} {arXiv preprint arXiv:2204.04623}\ } (\bibinfo {year}
  {2022})}\BibitemShut {NoStop}%
\bibitem [{\citenamefont {Semenok}\ \emph {et~al.}(2022)\citenamefont
  {Semenok}, \citenamefont {Troyan}, \citenamefont {Sadakov}, \citenamefont
  {Zhou}, \citenamefont {Galasso}, \citenamefont {Kvashnin}, \citenamefont
  {Ivanova}, \citenamefont {Kruglov}, \citenamefont {Bykov}, \citenamefont
  {Terent'ev} \emph {et~al.}}]{semenok2022effect}%
  \BibitemOpen
  \bibfield  {author} {\bibinfo {author} {\bibfnamefont {D.~V.}\ \bibnamefont
  {Semenok}}, \bibinfo {author} {\bibfnamefont {I.~A.}\ \bibnamefont {Troyan}},
  \bibinfo {author} {\bibfnamefont {A.~V.}\ \bibnamefont {Sadakov}}, \bibinfo
  {author} {\bibfnamefont {D.}~\bibnamefont {Zhou}}, \bibinfo {author}
  {\bibfnamefont {M.}~\bibnamefont {Galasso}}, \bibinfo {author} {\bibfnamefont
  {A.~G.}\ \bibnamefont {Kvashnin}}, \bibinfo {author} {\bibfnamefont {A.~G.}\
  \bibnamefont {Ivanova}}, \bibinfo {author} {\bibfnamefont {I.~A.}\
  \bibnamefont {Kruglov}}, \bibinfo {author} {\bibfnamefont {A.~A.}\
  \bibnamefont {Bykov}}, \bibinfo {author} {\bibfnamefont {K.~Y.}\ \bibnamefont
  {Terent'ev}},  \emph {et~al.},\ }\href@noop {} {\bibfield  {journal}
  {\bibinfo  {journal} {Adv. Mater.}\ }\textbf {\bibinfo {volume} {34}},\
  \bibinfo {pages} {2204038} (\bibinfo {year} {2022})}\BibitemShut {NoStop}%
\bibitem [{\citenamefont {Chen}\ \emph {et~al.}(2024)\citenamefont {Chen},
  \citenamefont {Qian}, \citenamefont {Huang}, \citenamefont {Chen},
  \citenamefont {Guo}, \citenamefont {Zhang}, \citenamefont {Zhang},
  \citenamefont {Yuan},\ and\ \citenamefont {Cui}}]{chen2024high}%
  \BibitemOpen
  \bibfield  {author} {\bibinfo {author} {\bibfnamefont {S.}~\bibnamefont
  {Chen}}, \bibinfo {author} {\bibfnamefont {Y.}~\bibnamefont {Qian}}, \bibinfo
  {author} {\bibfnamefont {X.}~\bibnamefont {Huang}}, \bibinfo {author}
  {\bibfnamefont {W.}~\bibnamefont {Chen}}, \bibinfo {author} {\bibfnamefont
  {J.}~\bibnamefont {Guo}}, \bibinfo {author} {\bibfnamefont {K.}~\bibnamefont
  {Zhang}}, \bibinfo {author} {\bibfnamefont {J.}~\bibnamefont {Zhang}},
  \bibinfo {author} {\bibfnamefont {H.}~\bibnamefont {Yuan}}, \ and\ \bibinfo
  {author} {\bibfnamefont {T.}~\bibnamefont {Cui}},\ }\href@noop {} {\bibfield
  {journal} {\bibinfo  {journal} {Nat. Sci. Rev.}\ }\textbf {\bibinfo {volume}
  {11}},\ \bibinfo {pages} {nwad107} (\bibinfo {year} {2024})}\BibitemShut
  {NoStop}%
\bibitem [{\citenamefont {Cui}\ \emph {et~al.}(2020)\citenamefont {Cui},
  \citenamefont {Bi}, \citenamefont {Shi}, \citenamefont {Li}, \citenamefont
  {Liu}, \citenamefont {Zurek},\ and\ \citenamefont {Hemley}}]{cui2020route}%
  \BibitemOpen
  \bibfield  {author} {\bibinfo {author} {\bibfnamefont {W.}~\bibnamefont
  {Cui}}, \bibinfo {author} {\bibfnamefont {T.}~\bibnamefont {Bi}}, \bibinfo
  {author} {\bibfnamefont {J.}~\bibnamefont {Shi}}, \bibinfo {author}
  {\bibfnamefont {Y.}~\bibnamefont {Li}}, \bibinfo {author} {\bibfnamefont
  {H.}~\bibnamefont {Liu}}, \bibinfo {author} {\bibfnamefont {E.}~\bibnamefont
  {Zurek}}, \ and\ \bibinfo {author} {\bibfnamefont {R.~J.}\ \bibnamefont
  {Hemley}},\ }\href@noop {} {\bibfield  {journal} {\bibinfo  {journal} {Phys.
  Rev. B}\ }\textbf {\bibinfo {volume} {101}},\ \bibinfo {pages} {134504}
  (\bibinfo {year} {2020})}\BibitemShut {NoStop}%
\bibitem [{\citenamefont {Jiang}\ \emph {et~al.}(2022)\citenamefont {Jiang},
  \citenamefont {Hai}, \citenamefont {Tian}, \citenamefont {Ding},
  \citenamefont {Feng}, \citenamefont {Yang}, \citenamefont {Chen},\ and\
  \citenamefont {Zhong}}]{jiang2022high}%
  \BibitemOpen
  \bibfield  {author} {\bibinfo {author} {\bibfnamefont {M.-J.}\ \bibnamefont
  {Jiang}}, \bibinfo {author} {\bibfnamefont {Y.-L.}\ \bibnamefont {Hai}},
  \bibinfo {author} {\bibfnamefont {H.-L.}\ \bibnamefont {Tian}}, \bibinfo
  {author} {\bibfnamefont {H.-B.}\ \bibnamefont {Ding}}, \bibinfo {author}
  {\bibfnamefont {Y.-J.}\ \bibnamefont {Feng}}, \bibinfo {author}
  {\bibfnamefont {C.-L.}\ \bibnamefont {Yang}}, \bibinfo {author}
  {\bibfnamefont {X.-J.}\ \bibnamefont {Chen}}, \ and\ \bibinfo {author}
  {\bibfnamefont {G.-H.}\ \bibnamefont {Zhong}},\ }\href@noop {} {\bibfield
  {journal} {\bibinfo  {journal} {Phys. Rev. B}\ }\textbf {\bibinfo {volume}
  {105}},\ \bibinfo {pages} {104511} (\bibinfo {year} {2022})}\BibitemShut
  {NoStop}%
\bibitem [{\citenamefont {Di~Cataldo}\ \emph {et~al.}(2021)\citenamefont
  {Di~Cataldo}, \citenamefont {Heil}, \citenamefont {von~der Linden},\ and\
  \citenamefont {Boeri}}]{di2021bh}%
  \BibitemOpen
  \bibfield  {author} {\bibinfo {author} {\bibfnamefont {S.}~\bibnamefont
  {Di~Cataldo}}, \bibinfo {author} {\bibfnamefont {C.}~\bibnamefont {Heil}},
  \bibinfo {author} {\bibfnamefont {W.}~\bibnamefont {von~der Linden}}, \ and\
  \bibinfo {author} {\bibfnamefont {L.}~\bibnamefont {Boeri}},\ }\href@noop {}
  {\bibfield  {journal} {\bibinfo  {journal} {Phys. Rev. B}\ }\textbf {\bibinfo
  {volume} {104}},\ \bibinfo {pages} {L020511} (\bibinfo {year}
  {2021})}\BibitemShut {NoStop}%
\bibitem [{\citenamefont {Liang}\ \emph {et~al.}(2021)\citenamefont {Liang},
  \citenamefont {Bergara}, \citenamefont {Wei}, \citenamefont {Song},
  \citenamefont {Wang}, \citenamefont {Sun}, \citenamefont {Liu}, \citenamefont
  {Hemley}, \citenamefont {Wang}, \citenamefont {Gao} \emph
  {et~al.}}]{liang2021prediction}%
  \BibitemOpen
  \bibfield  {author} {\bibinfo {author} {\bibfnamefont {X.}~\bibnamefont
  {Liang}}, \bibinfo {author} {\bibfnamefont {A.}~\bibnamefont {Bergara}},
  \bibinfo {author} {\bibfnamefont {X.}~\bibnamefont {Wei}}, \bibinfo {author}
  {\bibfnamefont {X.}~\bibnamefont {Song}}, \bibinfo {author} {\bibfnamefont
  {L.}~\bibnamefont {Wang}}, \bibinfo {author} {\bibfnamefont {R.}~\bibnamefont
  {Sun}}, \bibinfo {author} {\bibfnamefont {H.}~\bibnamefont {Liu}}, \bibinfo
  {author} {\bibfnamefont {R.~J.}\ \bibnamefont {Hemley}}, \bibinfo {author}
  {\bibfnamefont {L.}~\bibnamefont {Wang}}, \bibinfo {author} {\bibfnamefont
  {G.}~\bibnamefont {Gao}},  \emph {et~al.},\ }\href@noop {} {\bibfield
  {journal} {\bibinfo  {journal} {Phys. Rev. B}\ }\textbf {\bibinfo {volume}
  {104}},\ \bibinfo {pages} {134501} (\bibinfo {year} {2021})}\BibitemShut
  {NoStop}%
\bibitem [{\citenamefont {Belli}\ and\ \citenamefont
  {Errea}(2022)}]{belli2022impact}%
  \BibitemOpen
  \bibfield  {author} {\bibinfo {author} {\bibfnamefont {F.}~\bibnamefont
  {Belli}}\ and\ \bibinfo {author} {\bibfnamefont {I.}~\bibnamefont {Errea}},\
  }\href@noop {} {\bibfield  {journal} {\bibinfo  {journal} {Phys. Rev. B}\
  }\textbf {\bibinfo {volume} {106}},\ \bibinfo {pages} {134509} (\bibinfo
  {year} {2022})}\BibitemShut {NoStop}%
\bibitem [{\citenamefont {Song}\ \emph {et~al.}(2023)\citenamefont {Song},
  \citenamefont {Bi}, \citenamefont {Nakamoto}, \citenamefont {Shimizu},
  \citenamefont {Liu}, \citenamefont {Zou}, \citenamefont {Liu}, \citenamefont
  {Wang},\ and\ \citenamefont {Ma}}]{song2023stoichiometric}%
  \BibitemOpen
  \bibfield  {author} {\bibinfo {author} {\bibfnamefont {Y.}~\bibnamefont
  {Song}}, \bibinfo {author} {\bibfnamefont {J.}~\bibnamefont {Bi}}, \bibinfo
  {author} {\bibfnamefont {Y.}~\bibnamefont {Nakamoto}}, \bibinfo {author}
  {\bibfnamefont {K.}~\bibnamefont {Shimizu}}, \bibinfo {author} {\bibfnamefont
  {H.}~\bibnamefont {Liu}}, \bibinfo {author} {\bibfnamefont {B.}~\bibnamefont
  {Zou}}, \bibinfo {author} {\bibfnamefont {G.}~\bibnamefont {Liu}}, \bibinfo
  {author} {\bibfnamefont {H.}~\bibnamefont {Wang}}, \ and\ \bibinfo {author}
  {\bibfnamefont {Y.}~\bibnamefont {Ma}},\ }\href@noop {} {\bibfield  {journal}
  {\bibinfo  {journal} {Phys. Rev. Lett.}\ }\textbf {\bibinfo {volume} {130}},\
  \bibinfo {pages} {266001} (\bibinfo {year} {2023})}\BibitemShut {NoStop}%
\bibitem [{\citenamefont {Gao}\ \emph {et~al.}(2024)\citenamefont {Gao},
  \citenamefont {Cui}, \citenamefont {Shi}, \citenamefont {Durajski},
  \citenamefont {Hao}, \citenamefont {Botti}, \citenamefont {Marques},\ and\
  \citenamefont {Li}}]{gao2024prediction}%
  \BibitemOpen
  \bibfield  {author} {\bibinfo {author} {\bibfnamefont {K.}~\bibnamefont
  {Gao}}, \bibinfo {author} {\bibfnamefont {W.}~\bibnamefont {Cui}}, \bibinfo
  {author} {\bibfnamefont {J.}~\bibnamefont {Shi}}, \bibinfo {author}
  {\bibfnamefont {A.~P.}\ \bibnamefont {Durajski}}, \bibinfo {author}
  {\bibfnamefont {J.}~\bibnamefont {Hao}}, \bibinfo {author} {\bibfnamefont
  {S.}~\bibnamefont {Botti}}, \bibinfo {author} {\bibfnamefont {M.~A.}\
  \bibnamefont {Marques}}, \ and\ \bibinfo {author} {\bibfnamefont
  {Y.}~\bibnamefont {Li}},\ }\href@noop {} {\bibfield  {journal} {\bibinfo
  {journal} {Phys. Rev. B}\ }\textbf {\bibinfo {volume} {109}},\ \bibinfo
  {pages} {014501} (\bibinfo {year} {2024})}\BibitemShut {NoStop}%
\bibitem [{\citenamefont {Gao}\ \emph {et~al.}(2021)\citenamefont {Gao},
  \citenamefont {Yan}, \citenamefont {Lu},\ and\ \citenamefont
  {Xiang}}]{gao2021phonon}%
  \BibitemOpen
  \bibfield  {author} {\bibinfo {author} {\bibfnamefont {M.}~\bibnamefont
  {Gao}}, \bibinfo {author} {\bibfnamefont {X.-W.}\ \bibnamefont {Yan}},
  \bibinfo {author} {\bibfnamefont {Z.-Y.}\ \bibnamefont {Lu}}, \ and\ \bibinfo
  {author} {\bibfnamefont {T.}~\bibnamefont {Xiang}},\ }\href@noop {}
  {\bibfield  {journal} {\bibinfo  {journal} {Phys. Rev. B}\ }\textbf {\bibinfo
  {volume} {104}},\ \bibinfo {pages} {L100504} (\bibinfo {year}
  {2021})}\BibitemShut {NoStop}%
\bibitem [{\citenamefont {Li}\ \emph {et~al.}(2022{\natexlab{b}})\citenamefont
  {Li}, \citenamefont {Wang}, \citenamefont {Sun}, \citenamefont {Lu},\ and\
  \citenamefont {Peng}}]{li2022superconductivity2}%
  \BibitemOpen
  \bibfield  {author} {\bibinfo {author} {\bibfnamefont {S.}~\bibnamefont
  {Li}}, \bibinfo {author} {\bibfnamefont {H.}~\bibnamefont {Wang}}, \bibinfo
  {author} {\bibfnamefont {W.}~\bibnamefont {Sun}}, \bibinfo {author}
  {\bibfnamefont {C.}~\bibnamefont {Lu}}, \ and\ \bibinfo {author}
  {\bibfnamefont {F.}~\bibnamefont {Peng}},\ }\href@noop {} {\bibfield
  {journal} {\bibinfo  {journal} {Phys. Rev. B}\ }\textbf {\bibinfo {volume}
  {105}},\ \bibinfo {pages} {224107} (\bibinfo {year}
  {2022}{\natexlab{b}})}\BibitemShut {NoStop}%
\bibitem [{\citenamefont {Sanna}\ \emph {et~al.}(2024)\citenamefont {Sanna},
  \citenamefont {Cerqueira}, \citenamefont {Fang}, \citenamefont {Errea},
  \citenamefont {Ludwig},\ and\ \citenamefont {Marques}}]{sanna2024prediction}%
  \BibitemOpen
  \bibfield  {author} {\bibinfo {author} {\bibfnamefont {A.}~\bibnamefont
  {Sanna}}, \bibinfo {author} {\bibfnamefont {T.~F.}\ \bibnamefont
  {Cerqueira}}, \bibinfo {author} {\bibfnamefont {Y.-W.}\ \bibnamefont {Fang}},
  \bibinfo {author} {\bibfnamefont {I.}~\bibnamefont {Errea}}, \bibinfo
  {author} {\bibfnamefont {A.}~\bibnamefont {Ludwig}}, \ and\ \bibinfo {author}
  {\bibfnamefont {M.~A.}\ \bibnamefont {Marques}},\ }\href@noop {} {\bibfield
  {journal} {\bibinfo  {journal} {NPJ Comput. Mater}\ }\textbf {\bibinfo
  {volume} {10}},\ \bibinfo {pages} {44} (\bibinfo {year} {2024})}\BibitemShut
  {NoStop}%
\bibitem [{\citenamefont {Dolui}\ \emph {et~al.}(2024)\citenamefont {Dolui},
  \citenamefont {Conway}, \citenamefont {Heil}, \citenamefont {Strobel},
  \citenamefont {Prasankumar},\ and\ \citenamefont
  {Pickard}}]{dolui2024feasible}%
  \BibitemOpen
  \bibfield  {author} {\bibinfo {author} {\bibfnamefont {K.}~\bibnamefont
  {Dolui}}, \bibinfo {author} {\bibfnamefont {L.~J.}\ \bibnamefont {Conway}},
  \bibinfo {author} {\bibfnamefont {C.}~\bibnamefont {Heil}}, \bibinfo {author}
  {\bibfnamefont {T.~A.}\ \bibnamefont {Strobel}}, \bibinfo {author}
  {\bibfnamefont {R.~P.}\ \bibnamefont {Prasankumar}}, \ and\ \bibinfo {author}
  {\bibfnamefont {C.~J.}\ \bibnamefont {Pickard}},\ }\href@noop {} {\bibfield
  {journal} {\bibinfo  {journal} {Phys. Rev. Lett.}\ }\textbf {\bibinfo
  {volume} {132}},\ \bibinfo {pages} {166001} (\bibinfo {year}
  {2024})}\BibitemShut {NoStop}%
\bibitem [{\citenamefont {Song}\ \emph {et~al.}(2024)\citenamefont {Song},
  \citenamefont {Hao}, \citenamefont {Wei}, \citenamefont {He}, \citenamefont
  {Liu}, \citenamefont {Ma}, \citenamefont {Liu}, \citenamefont {Wang},
  \citenamefont {Niu}, \citenamefont {Wang} \emph
  {et~al.}}]{song2024superconductivity}%
  \BibitemOpen
  \bibfield  {author} {\bibinfo {author} {\bibfnamefont {X.}~\bibnamefont
  {Song}}, \bibinfo {author} {\bibfnamefont {X.}~\bibnamefont {Hao}}, \bibinfo
  {author} {\bibfnamefont {X.}~\bibnamefont {Wei}}, \bibinfo {author}
  {\bibfnamefont {X.-L.}\ \bibnamefont {He}}, \bibinfo {author} {\bibfnamefont
  {H.}~\bibnamefont {Liu}}, \bibinfo {author} {\bibfnamefont {L.}~\bibnamefont
  {Ma}}, \bibinfo {author} {\bibfnamefont {G.}~\bibnamefont {Liu}}, \bibinfo
  {author} {\bibfnamefont {H.}~\bibnamefont {Wang}}, \bibinfo {author}
  {\bibfnamefont {J.}~\bibnamefont {Niu}}, \bibinfo {author} {\bibfnamefont
  {S.}~\bibnamefont {Wang}},  \emph {et~al.},\ }\href@noop {} {\bibfield
  {journal} {\bibinfo  {journal} {J. Am. Chem. Soc}\ } (\bibinfo {year}
  {2024})}\BibitemShut {NoStop}%
\bibitem [{\citenamefont {Semenok}\ \emph {et~al.}(2018)\citenamefont
  {Semenok}, \citenamefont {Kvashnin}, \citenamefont {Kruglov},\ and\
  \citenamefont {Oganov}}]{semenok2018actinium}%
  \BibitemOpen
  \bibfield  {author} {\bibinfo {author} {\bibfnamefont {D.~V.}\ \bibnamefont
  {Semenok}}, \bibinfo {author} {\bibfnamefont {A.~G.}\ \bibnamefont
  {Kvashnin}}, \bibinfo {author} {\bibfnamefont {I.~A.}\ \bibnamefont
  {Kruglov}}, \ and\ \bibinfo {author} {\bibfnamefont {A.~R.}\ \bibnamefont
  {Oganov}},\ }\href@noop {} {\bibfield  {journal} {\bibinfo  {journal} {J.
  Phys. Chem. Lett}\ }\textbf {\bibinfo {volume} {9}},\ \bibinfo {pages} {1920}
  (\bibinfo {year} {2018})}\BibitemShut {NoStop}%
\bibitem [{\citenamefont {Kokail}\ \emph {et~al.}(2017)\citenamefont {Kokail},
  \citenamefont {Von Der~Linden},\ and\ \citenamefont
  {Boeri}}]{kokail2017prediction}%
  \BibitemOpen
  \bibfield  {author} {\bibinfo {author} {\bibfnamefont {C.}~\bibnamefont
  {Kokail}}, \bibinfo {author} {\bibfnamefont {W.}~\bibnamefont {Von
  Der~Linden}}, \ and\ \bibinfo {author} {\bibfnamefont {L.}~\bibnamefont
  {Boeri}},\ }\href@noop {} {\bibfield  {journal} {\bibinfo  {journal} {Phys.
  Rev. Mater}\ }\textbf {\bibinfo {volume} {1}},\ \bibinfo {pages} {074803}
  (\bibinfo {year} {2017})}\BibitemShut {NoStop}%
\bibitem [{\citenamefont {Du}\ \emph {et~al.}(2019)\citenamefont {Du},
  \citenamefont {Zhang}, \citenamefont {Lin}, \citenamefont {Zhang},
  \citenamefont {Bergara},\ and\ \citenamefont {Yang}}]{du2019phase}%
  \BibitemOpen
  \bibfield  {author} {\bibinfo {author} {\bibfnamefont {X.}~\bibnamefont
  {Du}}, \bibinfo {author} {\bibfnamefont {S.}~\bibnamefont {Zhang}}, \bibinfo
  {author} {\bibfnamefont {J.}~\bibnamefont {Lin}}, \bibinfo {author}
  {\bibfnamefont {X.}~\bibnamefont {Zhang}}, \bibinfo {author} {\bibfnamefont
  {A.}~\bibnamefont {Bergara}}, \ and\ \bibinfo {author} {\bibfnamefont
  {G.}~\bibnamefont {Yang}},\ }\href@noop {} {\bibfield  {journal} {\bibinfo
  {journal} {Phys. Rev. B}\ }\textbf {\bibinfo {volume} {100}},\ \bibinfo
  {pages} {134110} (\bibinfo {year} {2019})}\BibitemShut {NoStop}%
\bibitem [{\citenamefont {Di~Cataldo}\ \emph {et~al.}(2020)\citenamefont
  {Di~Cataldo}, \citenamefont {Von Der~Linden},\ and\ \citenamefont
  {Boeri}}]{di2020phase}%
  \BibitemOpen
  \bibfield  {author} {\bibinfo {author} {\bibfnamefont {S.}~\bibnamefont
  {Di~Cataldo}}, \bibinfo {author} {\bibfnamefont {W.}~\bibnamefont {Von
  Der~Linden}}, \ and\ \bibinfo {author} {\bibfnamefont {L.}~\bibnamefont
  {Boeri}},\ }\href@noop {} {\bibfield  {journal} {\bibinfo  {journal} {Phys.
  Rev. B}\ }\textbf {\bibinfo {volume} {102}},\ \bibinfo {pages} {014516}
  (\bibinfo {year} {2020})}\BibitemShut {NoStop}%
\bibitem [{\citenamefont {Li}\ \emph {et~al.}(2022{\natexlab{c}})\citenamefont
  {Li}, \citenamefont {Zhang}, \citenamefont {Bergara}, \citenamefont {Liu},\
  and\ \citenamefont {Yang}}]{li2022structural}%
  \BibitemOpen
  \bibfield  {author} {\bibinfo {author} {\bibfnamefont {X.}~\bibnamefont
  {Li}}, \bibinfo {author} {\bibfnamefont {X.}~\bibnamefont {Zhang}}, \bibinfo
  {author} {\bibfnamefont {A.}~\bibnamefont {Bergara}}, \bibinfo {author}
  {\bibfnamefont {Y.}~\bibnamefont {Liu}}, \ and\ \bibinfo {author}
  {\bibfnamefont {G.}~\bibnamefont {Yang}},\ }\href@noop {} {\bibfield
  {journal} {\bibinfo  {journal} {Phys. Rev. B}\ }\textbf {\bibinfo {volume}
  {106}},\ \bibinfo {pages} {174104} (\bibinfo {year}
  {2022}{\natexlab{c}})}\BibitemShut {NoStop}%
\bibitem [{\citenamefont {Sukmas}\ \emph {et~al.}(2023)\citenamefont {Sukmas},
  \citenamefont {Tsuppayakorn-aek}, \citenamefont {Pluengphon}, \citenamefont
  {Clark}, \citenamefont {Ahuja}, \citenamefont {Bovornratanaraks},\ and\
  \citenamefont {Luo}}]{sukmas2023first}%
  \BibitemOpen
  \bibfield  {author} {\bibinfo {author} {\bibfnamefont {W.}~\bibnamefont
  {Sukmas}}, \bibinfo {author} {\bibfnamefont {P.}~\bibnamefont
  {Tsuppayakorn-aek}}, \bibinfo {author} {\bibfnamefont {P.}~\bibnamefont
  {Pluengphon}}, \bibinfo {author} {\bibfnamefont {S.~J.}\ \bibnamefont
  {Clark}}, \bibinfo {author} {\bibfnamefont {R.}~\bibnamefont {Ahuja}},
  \bibinfo {author} {\bibfnamefont {T.}~\bibnamefont {Bovornratanaraks}}, \
  and\ \bibinfo {author} {\bibfnamefont {W.}~\bibnamefont {Luo}},\ }\href@noop
  {} {\bibfield  {journal} {\bibinfo  {journal} {INT J HYDROGEN ENERG}\
  }\textbf {\bibinfo {volume} {48}},\ \bibinfo {pages} {4006} (\bibinfo {year}
  {2023})}\BibitemShut {NoStop}%
\bibitem [{\citenamefont {Wang}\ \emph {et~al.}(2010)\citenamefont {Wang},
  \citenamefont {Lv}, \citenamefont {Zhu},\ and\ \citenamefont
  {Ma}}]{wang2010crystal}%
  \BibitemOpen
  \bibfield  {author} {\bibinfo {author} {\bibfnamefont {Y.}~\bibnamefont
  {Wang}}, \bibinfo {author} {\bibfnamefont {J.}~\bibnamefont {Lv}}, \bibinfo
  {author} {\bibfnamefont {L.}~\bibnamefont {Zhu}}, \ and\ \bibinfo {author}
  {\bibfnamefont {Y.}~\bibnamefont {Ma}},\ }\href@noop {} {\bibfield  {journal}
  {\bibinfo  {journal} {Phys. Rev. B}\ }\textbf {\bibinfo {volume} {82}},\
  \bibinfo {pages} {094116} (\bibinfo {year} {2010})}\BibitemShut {NoStop}%
\bibitem [{\citenamefont {Wang}\ \emph
  {et~al.}(2012{\natexlab{b}})\citenamefont {Wang}, \citenamefont {Lv},
  \citenamefont {Zhu},\ and\ \citenamefont {Ma}}]{wang2012calypso}%
  \BibitemOpen
  \bibfield  {author} {\bibinfo {author} {\bibfnamefont {Y.}~\bibnamefont
  {Wang}}, \bibinfo {author} {\bibfnamefont {J.}~\bibnamefont {Lv}}, \bibinfo
  {author} {\bibfnamefont {L.}~\bibnamefont {Zhu}}, \ and\ \bibinfo {author}
  {\bibfnamefont {Y.}~\bibnamefont {Ma}},\ }\href@noop {} {\bibfield  {journal}
  {\bibinfo  {journal} {Comput. Phys. Commun.}\ }\textbf {\bibinfo {volume}
  {183}},\ \bibinfo {pages} {2063} (\bibinfo {year}
  {2012}{\natexlab{b}})}\BibitemShut {NoStop}%
\bibitem [{\citenamefont {Gao}\ \emph {et~al.}(2019)\citenamefont {Gao},
  \citenamefont {Gao}, \citenamefont {Lu}, \citenamefont {Lv}, \citenamefont
  {Wang},\ and\ \citenamefont {Ma}}]{gao2019interface}%
  \BibitemOpen
  \bibfield  {author} {\bibinfo {author} {\bibfnamefont {B.}~\bibnamefont
  {Gao}}, \bibinfo {author} {\bibfnamefont {P.}~\bibnamefont {Gao}}, \bibinfo
  {author} {\bibfnamefont {S.}~\bibnamefont {Lu}}, \bibinfo {author}
  {\bibfnamefont {J.}~\bibnamefont {Lv}}, \bibinfo {author} {\bibfnamefont
  {Y.}~\bibnamefont {Wang}}, \ and\ \bibinfo {author} {\bibfnamefont
  {Y.}~\bibnamefont {Ma}},\ }\href@noop {} {\bibfield  {journal} {\bibinfo
  {journal} {Sci. Bull.}\ }\textbf {\bibinfo {volume} {64}},\ \bibinfo {pages}
  {301} (\bibinfo {year} {2019})}\BibitemShut {NoStop}%
\bibitem [{\citenamefont {Shao}\ \emph {et~al.}(2022)\citenamefont {Shao},
  \citenamefont {Lv}, \citenamefont {Liu}, \citenamefont {Shao}, \citenamefont
  {Gao}, \citenamefont {Liu}, \citenamefont {Wang},\ and\ \citenamefont
  {Ma}}]{shao2022symmetry}%
  \BibitemOpen
  \bibfield  {author} {\bibinfo {author} {\bibfnamefont {X.}~\bibnamefont
  {Shao}}, \bibinfo {author} {\bibfnamefont {J.}~\bibnamefont {Lv}}, \bibinfo
  {author} {\bibfnamefont {P.}~\bibnamefont {Liu}}, \bibinfo {author}
  {\bibfnamefont {S.}~\bibnamefont {Shao}}, \bibinfo {author} {\bibfnamefont
  {P.}~\bibnamefont {Gao}}, \bibinfo {author} {\bibfnamefont {H.}~\bibnamefont
  {Liu}}, \bibinfo {author} {\bibfnamefont {Y.}~\bibnamefont {Wang}}, \ and\
  \bibinfo {author} {\bibfnamefont {Y.}~\bibnamefont {Ma}},\ }\href@noop {}
  {\bibfield  {journal} {\bibinfo  {journal} {J. Chem. Phys.}\ }\textbf
  {\bibinfo {volume} {156}},\ \bibinfo {pages} {014105} (\bibinfo {year}
  {2022})}\BibitemShut {NoStop}%
\bibitem [{\citenamefont {Parlinski}\ \emph {et~al.}(1997)\citenamefont
  {Parlinski}, \citenamefont {Li},\ and\ \citenamefont
  {Kawazoe}}]{parlinski1997first}%
  \BibitemOpen
  \bibfield  {author} {\bibinfo {author} {\bibfnamefont {K.}~\bibnamefont
  {Parlinski}}, \bibinfo {author} {\bibfnamefont {Z.}~\bibnamefont {Li}}, \
  and\ \bibinfo {author} {\bibfnamefont {Y.}~\bibnamefont {Kawazoe}},\
  }\href@noop {} {\bibfield  {journal} {\bibinfo  {journal} {Phys. Rev. Lett.}\
  }\textbf {\bibinfo {volume} {78}},\ \bibinfo {pages} {4063} (\bibinfo {year}
  {1997})}\BibitemShut {NoStop}%
\bibitem [{\citenamefont {Togo}\ \emph {et~al.}(2008)\citenamefont {Togo},
  \citenamefont {Oba},\ and\ \citenamefont {Tanaka}}]{togo2008first}%
  \BibitemOpen
  \bibfield  {author} {\bibinfo {author} {\bibfnamefont {A.}~\bibnamefont
  {Togo}}, \bibinfo {author} {\bibfnamefont {F.}~\bibnamefont {Oba}}, \ and\
  \bibinfo {author} {\bibfnamefont {I.}~\bibnamefont {Tanaka}},\ }\href@noop {}
  {\bibfield  {journal} {\bibinfo  {journal} {Phys. Rev. B}\ }\textbf {\bibinfo
  {volume} {78}},\ \bibinfo {pages} {134106} (\bibinfo {year}
  {2008})}\BibitemShut {NoStop}%
\bibitem [{\citenamefont {Kresse}\ and\ \citenamefont
  {Furthm{\"u}ller}(1996)}]{kresse1996efficient}%
  \BibitemOpen
  \bibfield  {author} {\bibinfo {author} {\bibfnamefont {G.}~\bibnamefont
  {Kresse}}\ and\ \bibinfo {author} {\bibfnamefont {J.}~\bibnamefont
  {Furthm{\"u}ller}},\ }\href@noop {} {\bibfield  {journal} {\bibinfo
  {journal} {Phys. Rev. B}\ }\textbf {\bibinfo {volume} {54}},\ \bibinfo
  {pages} {11169} (\bibinfo {year} {1996})}\BibitemShut {NoStop}%
\bibitem [{\citenamefont {Perdew}\ \emph {et~al.}(1996)\citenamefont {Perdew},
  \citenamefont {Burke},\ and\ \citenamefont
  {Ernzerhof}}]{perdew1996generalized}%
  \BibitemOpen
  \bibfield  {author} {\bibinfo {author} {\bibfnamefont {J.~P.}\ \bibnamefont
  {Perdew}}, \bibinfo {author} {\bibfnamefont {K.}~\bibnamefont {Burke}}, \
  and\ \bibinfo {author} {\bibfnamefont {M.}~\bibnamefont {Ernzerhof}},\
  }\href@noop {} {\bibfield  {journal} {\bibinfo  {journal} {Phys. Rev. Lett.}\
  }\textbf {\bibinfo {volume} {77}},\ \bibinfo {pages} {3865} (\bibinfo {year}
  {1996})}\BibitemShut {NoStop}%
\bibitem [{\citenamefont {Perdew}\ \emph {et~al.}(1992)\citenamefont {Perdew},
  \citenamefont {Chevary}, \citenamefont {Vosko}, \citenamefont {Jackson},
  \citenamefont {Pederson}, \citenamefont {Singh},\ and\ \citenamefont
  {Fiolhais}}]{perdew1992atoms}%
  \BibitemOpen
  \bibfield  {author} {\bibinfo {author} {\bibfnamefont {J.~P.}\ \bibnamefont
  {Perdew}}, \bibinfo {author} {\bibfnamefont {J.~A.}\ \bibnamefont {Chevary}},
  \bibinfo {author} {\bibfnamefont {S.~H.}\ \bibnamefont {Vosko}}, \bibinfo
  {author} {\bibfnamefont {K.~A.}\ \bibnamefont {Jackson}}, \bibinfo {author}
  {\bibfnamefont {M.~R.}\ \bibnamefont {Pederson}}, \bibinfo {author}
  {\bibfnamefont {D.~J.}\ \bibnamefont {Singh}}, \ and\ \bibinfo {author}
  {\bibfnamefont {C.}~\bibnamefont {Fiolhais}},\ }\href@noop {} {\bibfield
  {journal} {\bibinfo  {journal} {Phys. Rev. B}\ }\textbf {\bibinfo {volume}
  {46}},\ \bibinfo {pages} {6671} (\bibinfo {year} {1992})}\BibitemShut
  {NoStop}%
\bibitem [{\citenamefont {Kresse}\ and\ \citenamefont
  {Joubert}(1999)}]{kresse1999ultrasoft}%
  \BibitemOpen
  \bibfield  {author} {\bibinfo {author} {\bibfnamefont {G.}~\bibnamefont
  {Kresse}}\ and\ \bibinfo {author} {\bibfnamefont {D.}~\bibnamefont
  {Joubert}},\ }\href@noop {} {\bibfield  {journal} {\bibinfo  {journal} {Phys.
  Rev. B}\ }\textbf {\bibinfo {volume} {59}},\ \bibinfo {pages} {1758}
  (\bibinfo {year} {1999})}\BibitemShut {NoStop}%
\bibitem [{\citenamefont {Becke}\ and\ \citenamefont
  {Edgecombe}(1990)}]{becke1990simple}%
  \BibitemOpen
  \bibfield  {author} {\bibinfo {author} {\bibfnamefont {A.~D.}\ \bibnamefont
  {Becke}}\ and\ \bibinfo {author} {\bibfnamefont {K.~E.}\ \bibnamefont
  {Edgecombe}},\ }\href@noop {} {\bibfield  {journal} {\bibinfo  {journal} {J.
  Chem. Phys.}\ }\textbf {\bibinfo {volume} {92}},\ \bibinfo {pages} {5397}
  (\bibinfo {year} {1990})}\BibitemShut {NoStop}%
\bibitem [{\citenamefont {Bader}(1985)}]{bader1985atoms}%
  \BibitemOpen
  \bibfield  {author} {\bibinfo {author} {\bibfnamefont {R.~F.}\ \bibnamefont
  {Bader}},\ }\href@noop {} {\bibfield  {journal} {\bibinfo  {journal} {Acc.
  Chem. Res.}\ }\textbf {\bibinfo {volume} {18}},\ \bibinfo {pages} {9}
  (\bibinfo {year} {1985})}\BibitemShut {NoStop}%
\bibitem [{\citenamefont {Giannozzi}\ \emph {et~al.}(2009)\citenamefont
  {Giannozzi}, \citenamefont {Baroni}, \citenamefont {Bonini}, \citenamefont
  {Calandra}, \citenamefont {Car}, \citenamefont {Cavazzoni}, \citenamefont
  {Ceresoli}, \citenamefont {Chiarotti}, \citenamefont {Cococcioni},
  \citenamefont {Dabo} \emph {et~al.}}]{giannozzi2009quantum}%
  \BibitemOpen
  \bibfield  {author} {\bibinfo {author} {\bibfnamefont {P.}~\bibnamefont
  {Giannozzi}}, \bibinfo {author} {\bibfnamefont {S.}~\bibnamefont {Baroni}},
  \bibinfo {author} {\bibfnamefont {N.}~\bibnamefont {Bonini}}, \bibinfo
  {author} {\bibfnamefont {M.}~\bibnamefont {Calandra}}, \bibinfo {author}
  {\bibfnamefont {R.}~\bibnamefont {Car}}, \bibinfo {author} {\bibfnamefont
  {C.}~\bibnamefont {Cavazzoni}}, \bibinfo {author} {\bibfnamefont
  {D.}~\bibnamefont {Ceresoli}}, \bibinfo {author} {\bibfnamefont {G.~L.}\
  \bibnamefont {Chiarotti}}, \bibinfo {author} {\bibfnamefont {M.}~\bibnamefont
  {Cococcioni}}, \bibinfo {author} {\bibfnamefont {I.}~\bibnamefont {Dabo}},
  \emph {et~al.},\ }\href@noop {} {\bibfield  {journal} {\bibinfo  {journal}
  {J. Phys.: Condens. Matter}\ }\textbf {\bibinfo {volume} {21}},\ \bibinfo
  {pages} {395502} (\bibinfo {year} {2009})}\BibitemShut {NoStop}%
\bibitem [{sup()}]{suppe}%
  \BibitemOpen
  \href@noop {} {}\bibinfo {note} {See Supplemental Material at [URL will be
  inserted by publisher]. It contains the charges transfer, the electronic
  properties and phonon dispersions of other compounds at different pressure
  and structrue information of all the compounds, etc.}\BibitemShut {Stop}%
\bibitem [{\citenamefont {Allen}\ and\ \citenamefont
  {Dynes}(1975)}]{allen1975transition}%
  \BibitemOpen
  \bibfield  {author} {\bibinfo {author} {\bibfnamefont {P.~B.}\ \bibnamefont
  {Allen}}\ and\ \bibinfo {author} {\bibfnamefont {R.}~\bibnamefont {Dynes}},\
  }\href@noop {} {\bibfield  {journal} {\bibinfo  {journal} {Phys. Rev. B}\
  }\textbf {\bibinfo {volume} {12}},\ \bibinfo {pages} {905} (\bibinfo {year}
  {1975})}\BibitemShut {NoStop}%
\bibitem [{\citenamefont {Yao}\ and\ \citenamefont
  {Hoffmann}(2011)}]{yao2011bh3}%
  \BibitemOpen
  \bibfield  {author} {\bibinfo {author} {\bibfnamefont {Y.}~\bibnamefont
  {Yao}}\ and\ \bibinfo {author} {\bibfnamefont {R.}~\bibnamefont {Hoffmann}},\
  }\href@noop {} {\bibfield  {journal} {\bibinfo  {journal} {J. Am. Chem. Soc}\
  }\textbf {\bibinfo {volume} {133}},\ \bibinfo {pages} {21002} (\bibinfo
  {year} {2011})}\BibitemShut {NoStop}%
\bibitem [{\citenamefont {Abe}\ and\ \citenamefont
  {Ashcroft}(2011)}]{abe2011crystalline}%
  \BibitemOpen
  \bibfield  {author} {\bibinfo {author} {\bibfnamefont {K.}~\bibnamefont
  {Abe}}\ and\ \bibinfo {author} {\bibfnamefont {N.}~\bibnamefont {Ashcroft}},\
  }\href@noop {} {\bibfield  {journal} {\bibinfo  {journal} {Phys. Rev. B}\
  }\textbf {\bibinfo {volume} {84}},\ \bibinfo {pages} {104118} (\bibinfo
  {year} {2011})}\BibitemShut {NoStop}%
\bibitem [{\citenamefont {Hu}\ \emph {et~al.}(2013)\citenamefont {Hu},
  \citenamefont {Oganov}, \citenamefont {Zhu}, \citenamefont {Qian},
  \citenamefont {Frapper}, \citenamefont {Lyakhov},\ and\ \citenamefont
  {Zhou}}]{hu2013pressure}%
  \BibitemOpen
  \bibfield  {author} {\bibinfo {author} {\bibfnamefont {C.-H.}\ \bibnamefont
  {Hu}}, \bibinfo {author} {\bibfnamefont {A.~R.}\ \bibnamefont {Oganov}},
  \bibinfo {author} {\bibfnamefont {Q.}~\bibnamefont {Zhu}}, \bibinfo {author}
  {\bibfnamefont {G.-R.}\ \bibnamefont {Qian}}, \bibinfo {author}
  {\bibfnamefont {G.}~\bibnamefont {Frapper}}, \bibinfo {author} {\bibfnamefont
  {A.~O.}\ \bibnamefont {Lyakhov}}, \ and\ \bibinfo {author} {\bibfnamefont
  {H.-Y.}\ \bibnamefont {Zhou}},\ }\href@noop {} {\bibfield  {journal}
  {\bibinfo  {journal} {Phys. Rev. Lett.}\ }\textbf {\bibinfo {volume} {110}},\
  \bibinfo {pages} {165504} (\bibinfo {year} {2013})}\BibitemShut {NoStop}%
\bibitem [{\citenamefont {Torabi}\ \emph {et~al.}(2013)\citenamefont {Torabi},
  \citenamefont {Song},\ and\ \citenamefont {Staroverov}}]{torabi2013pressure}%
  \BibitemOpen
  \bibfield  {author} {\bibinfo {author} {\bibfnamefont {A.}~\bibnamefont
  {Torabi}}, \bibinfo {author} {\bibfnamefont {Y.}~\bibnamefont {Song}}, \ and\
  \bibinfo {author} {\bibfnamefont {V.~N.}\ \bibnamefont {Staroverov}},\
  }\href@noop {} {\bibfield  {journal} {\bibinfo  {journal} {J. Phys. Chem. C}\
  }\textbf {\bibinfo {volume} {117}},\ \bibinfo {pages} {2210} (\bibinfo {year}
  {2013})}\BibitemShut {NoStop}%
\bibitem [{\citenamefont {Murcia~Rios}\ \emph {et~al.}(2018)\citenamefont
  {Murcia~Rios}, \citenamefont {Komsa},\ and\ \citenamefont
  {Staroverov}}]{murcia2018effects}%
  \BibitemOpen
  \bibfield  {author} {\bibinfo {author} {\bibfnamefont {A.~M.}\ \bibnamefont
  {Murcia~Rios}}, \bibinfo {author} {\bibfnamefont {D.~N.}\ \bibnamefont
  {Komsa}}, \ and\ \bibinfo {author} {\bibfnamefont {V.~N.}\ \bibnamefont
  {Staroverov}},\ }\href@noop {} {\bibfield  {journal} {\bibinfo  {journal} {J.
  Phys. Chem. C}\ }\textbf {\bibinfo {volume} {122}},\ \bibinfo {pages} {14781}
  (\bibinfo {year} {2018})}\BibitemShut {NoStop}%
\bibitem [{\citenamefont {Yang}\ \emph {et~al.}(2019)\citenamefont {Yang},
  \citenamefont {Lu}, \citenamefont {Li}, \citenamefont {Xue}, \citenamefont
  {Zang}, \citenamefont {Ho},\ and\ \citenamefont {Wang}}]{yang2019novel}%
  \BibitemOpen
  \bibfield  {author} {\bibinfo {author} {\bibfnamefont {W.-H.}\ \bibnamefont
  {Yang}}, \bibinfo {author} {\bibfnamefont {W.-C.}\ \bibnamefont {Lu}},
  \bibinfo {author} {\bibfnamefont {S.-D.}\ \bibnamefont {Li}}, \bibinfo
  {author} {\bibfnamefont {X.-Y.}\ \bibnamefont {Xue}}, \bibinfo {author}
  {\bibfnamefont {Q.-J.}\ \bibnamefont {Zang}}, \bibinfo {author}
  {\bibfnamefont {K.-M.}\ \bibnamefont {Ho}}, \ and\ \bibinfo {author}
  {\bibfnamefont {C.-Z.}\ \bibnamefont {Wang}},\ }\href@noop {} {\bibfield
  {journal} {\bibinfo  {journal} {Phys. Chem. Chem. Phys}\ }\textbf {\bibinfo
  {volume} {21}},\ \bibinfo {pages} {5466} (\bibinfo {year}
  {2019})}\BibitemShut {NoStop}%
\bibitem [{\citenamefont {Li}\ \emph {et~al.}(2023)\citenamefont {Li},
  \citenamefont {Yang},\ and\ \citenamefont {Lu}}]{li2023pressure}%
  \BibitemOpen
  \bibfield  {author} {\bibinfo {author} {\bibfnamefont {W.-H.}\ \bibnamefont
  {Li}}, \bibinfo {author} {\bibfnamefont {W.-H.}\ \bibnamefont {Yang}}, \ and\
  \bibinfo {author} {\bibfnamefont {W.-C.}\ \bibnamefont {Lu}},\ }\href@noop {}
  {\bibfield  {journal} {\bibinfo  {journal} {Phys. Chem. Chem. Phys}\ }\textbf
  {\bibinfo {volume} {25}},\ \bibinfo {pages} {22032} (\bibinfo {year}
  {2023})}\BibitemShut {NoStop}%
\bibitem [{\citenamefont {Eliashberg}(1960)}]{eliashberg1960interactions}%
  \BibitemOpen
  \bibfield  {author} {\bibinfo {author} {\bibfnamefont {G.}~\bibnamefont
  {Eliashberg}},\ }\href@noop {} {\bibfield  {journal} {\bibinfo  {journal}
  {Sov. Phys. JETP}\ }\textbf {\bibinfo {volume} {11}},\ \bibinfo {pages} {696}
  (\bibinfo {year} {1960})}\BibitemShut {NoStop}%
\end{thebibliography}%

\end{document}